\documentclass[journal]{IEEEtran}
\usepackage{cite}

\newcommand{\wvec}{{\bf{w}}}
\newcommand{\xvec}{{\bf{x}}}

\newcommand{\zerovec}{{\bf{0}}}

\newcommand{\Amat}{{\bf{A}}}
\newcommand{\Bmat}{{\bf{B}}}

\newcommand{\Hmat}{{\bf{H}}}

\newcommand{\Imat}{{\bf{I}}}

\newcommand{\Pmat}{{\bf{P}}}

\newcommand{\ind}[1]{\mathbbm{1}_{\{#1\}}}

\newcommand{\define}{\stackrel{\triangle}{=}}

\def\bsigma{{\mbox{\boldmath $\Sigma$}}}

\def\bvarphi{{\mbox{\boldmath $\varphi$}}}
\def\bvarphismall{{\mbox{\boldmath {\scriptsize $\varphi$}}}}

\def\thetavec{{\mbox{\boldmath $\theta$}}}

\def\muvec{{\mbox{\boldmath $\mu$}}}

\newcommand{\be}{\begin{equation}}
\newcommand{\ee}{\end{equation}}
\newcommand{\beqna}{\begin{eqnarray}}
\newcommand{\eeqna}{\end{eqnarray}}

\usepackage{verbatim}

\usepackage{amsfonts,amssymb,amsmath,bbm,mathtools,stackrel,amsthm} 
\usepackage{enumitem}
\usepackage{xr-hyper}
\usepackage[hidelinks]{hyperref}
\usepackage{textcase}
\usepackage{graphicx,subcaption}
\usepackage{epsfig,epstopdf}

\graphicspath{{figs/}} 
\usepackage{placeins}  
\usepackage{tikz}

\usetikzlibrary{shapes,arrows,fit,positioning}
\usetikzlibrary{calc}
\usetikzlibrary{decorations.markings}
\tikzstyle{block} = [draw, fill=white, rectangle, 
minimum height=3em, minimum width=6em]

\tikzstyle{input} = [anchor=west]
\tikzstyle{output} = [anchor=east]
\tikzstyle{pinstyle} = [pin edge={to-,thin,black}]
\usetikzlibrary{fit,tikzmark}

\usepackage[capitalise,noabbrev]{cleveref}

\crefname{subsection}{Subsection}{Subsections}
\Crefname{appendix}{Appendix}{Appendices}
\Crefname{objective}{Objective}{Objectives}
\Crefname{condition}{Condition}{Conditions}
\Crefname{reg_condition}{regularity Condition}{regularity Conditions}
\crefname{Theorem}{Theorem}{Theorems}
\Crefname{equation}{}{}
\crefname{figure}{Fig.}{Figs.}

\newcommand \grad[1]{\nabla_{\hspace{-0.1cm}{\scriptscriptstyle#1}}} 
\newcommand \der[2]{\frac{\partial  #1}{\partial #2}} 
 
\newcommand \eval[1] {\raisebox{-.4em}{$\Bigr\rvert_{\scriptstyle#1} $}}
\newcommand\norm[1]{\left\lVert#1\right\rVert}
\newcommand \Am[1]{{\cal A}_#1}
 
\newcommand \kld[2]{{\cal D}_{KL} \left( #1||#2   \right)}

\newcommand \oml[1] {\hat{#1}_{ \text{\tiny{OML}}}}
\newcommand \msl[1] {\hat{#1}_{ \text{\tiny{MSL} }}}
\newcommand \msnl[1] {\hat{#1}_{ \text{\tiny{ MSNL} }}}
\newcommand \mml[1] {\hat{#1}_{\text{\tiny{ MML} }}}
\newcommand \psml[1] {\hat{#1}_{\text{\tiny{ PSML}}}}

\def \thvec {\boldsymbol \theta}
\def \thvecvar {\boldsymbol \vartheta}

\newtheorem{Theorem}{Theorem}
\newtheorem{remark}{Remark}

\newtheorem{definition}{Definition}

\newtheorem{claim}{Claim}

\title{Non-Bayesian Post-Model-Selection Estimation as Estimation Under Model Misspecification}
\author{Nadav~Harel,~\IEEEmembership{Student Member,~IEEE},
	and~Tirza~Routtenberg,~\IEEEmembership{Senior Member,~IEEE} \vspace{-0.5cm}
	\thanks{N. Harel and T. Routtenberg are with the School of Electrical and Computer Engineering (ECE), Ben-Gurion University of the Negev,
		Beer-Sheva 84105, Israel. T. Routtenberg is also with the Department of ECE, Princeton University, Princeton, NJ.   e-mail: nadavhar@post.bgu.ac.il,~tirzar@bgu.ac.il.
		This research was partially supported by the 
		ISRAEL SCIENCE FOUNDATION (Grant No. 1148/22) and by the Pazi Foundation.
		Nadav Harel has been funded by the Kreitman School of Advanced Graduate
		Studies.
	}
}

\begin{document} \bstctlcite{IEEEexample:BSTcontrol}
	\maketitle
	\begin{abstract}
			In many parameter estimation problems, the exact model is unknown and is assumed to belong to a set of candidate models. In such cases, a predetermined data-based selection rule selects a  parametric model from a set of candidates before the parameter estimation.  
 The existing framework for  estimation under model misspecification   does not account for the selection process that led to the misspecified model. Moreover, in post-model-selection estimation, there are multiple candidate models chosen based on the observations, making the interpretation of the assumed model in the misspecified setting non-trivial. In this work, we present three interpretations to address the problem of non-Bayesian post-model-selection estimation as an estimation under model misspecification problem: the naive interpretation, the normalized interpretation, and the selective inference interpretation, and discuss their properties. For each of these interpretations, we developed the corresponding misspecified maximum likelihood estimator and the misspecified Cram{$\acute{\text{e}}$}r-Rao-type lower bound. The relations between the estimators and the performance bounds, as well as their properties, are discussed. Finally, we demonstrate the performance of the proposed estimators and bounds
 via simulations of estimation after channel selection. We show that the proposed performance bounds are more informative than the oracle Cram{$\acute{\text{e}}$}r-Rao Bound (CRB), where the third interpretation (selective inference) results in the lowest mean-squared-error (MSE) among the estimators.  
	\end{abstract}
	\section{Introduction}
	

	Traditional estimation methods and performance analysis are based on the assumption that the observation model is correctly specified. However, this assumption does not hold in many applications of signal processing, communication, and data science. 
	In many practical parameter estimation problems, the exact model is unknown and is assumed to belong to a set of candidate models.
	Therefore, before estimation, a model-selection stage is performed 
 by a predetermined data-based selection rule, where the selected model may be different from the true one, which affects the consequent estimation approach. This problem arises, for example,  in 
 direction of arrival (DOA) estimation where there is a detection stage prior to the DOA estimation \cite{chaumette2005influence} and in post-model-selection estimation of rain level from 
 commercial microwave links \cite{weiss2021total}.
	
Bounds and estimation methods have been developed for post-{\em parameter}-selection estimation \cite{routtenberg2016estimation,harel2020low,harel2021post}, where 
a subset of parameters of interest is selected, based on the data, prior to the estimation. In \cite{bashan2008optimal,bazerque2010distributed,msechu2012sensor}, the bounds and estimators were developed for cases where the informative data region is selected prior to the estimation. 
Post-model-selection estimation has been discussed as a key component of the framework of selective inference. 
	In \cite{shen2004inference,berk2013valid,efron2014estimation,leeb2005model,lee2016exact,fithian2014optimal,zollner2007overcoming,heller2019post,belloni2013least}, it has been shown that ignoring the model-selection procedure may cause overoptimistic inferences, non-covering confidence intervals, and introduces selection bias. In the context of signal processing, \cite{chaumette2005influence,chaumette2007cramer} presented the effect of a preliminary detection step on the estimation procedure.
	In \cite{meir2021cramer,harel2021bayesian} post-model-selection Cram$\acute{\text{e}}$r-Rao-type lower bounds on the mean-squared-error (MSE) were developed for non-Bayesian and Bayesian parameter estimation, respectively. 
 However, these bounds only fit
 nested candidate models, and they have not been formulated in terms of misspecification. 

	Estimation under misspecified models has been discussed in the literature and garnered renewed attention in recent years.
	In particular, \cite{huber1967under,akaike1998information,white1982maximum} discuss the asymptotic properties of the maximum likelihood (ML) estimator under a misspecified model, also known as the misspecified ML (MML) estimator.  For Bayesian parameter estimation, \cite{berk1966limiting,bunke1998asymptotic} discuss the properties of Bayesian estimators under misspecified models. A Cram$\acute{\text{e}}$r-Rao bound (CRB) that accounts for model misspecification, the misspecified CRB (MCRB), was developed in \cite{vuong1986cramer,richmond2015parameter} and was discussed in and derived for several scenarios in  \cite{fortunati2016constrained,fortunati2016misspecified,fortunati2017performance,fortunati2019semiparametric,abed2021misspecified,levy2023mcrb}. 
 In \cite{krauz2021composite,rosenthal2022model}, the MCRB was used to design a model-selection procedure. In \cite{weiss2023bilateral}, a bilateral bound on the MSE under model mismatch that is applicable for Bayesian and non-Bayesian approaches were developed.
	Despite the elegant and useful theory presented in these works, none of the existing works deals with post-model-selection estimation and considers the procedure that led to the selection of the misspecified model.
 In the	conventional estimation under model misspecification, there is a clear definition of the assumed probability density function (pdf). However, in post-model-selection estimation, there are several candidate models. Therefore, the interpretation of the assumed model, in this case, is not straightforward. As a result, the existing misspecified bounds and estimators cannot be used for the considered post-model-selection estimation.

	In this paper, we address the problem of estimation after model selection via the theory and methods developed for estimation under model misspecification. To this end, we present three interpretations of post-model-selection estimation as an estimation under model misspecification problem: 1) the commonly-used naive interpretation that defines the assumed pdf as the pdf of the selected model, which results in a non-valid pdf; 2)  the normalized interpretation, which is obtained by normalizing the naive pdf to a valid pdf, but creates coupling between the selected and unselected parameters; 3) selective inference interpretation, which is both valid (normalized) and without an unnecessary coupling between the parameters, and is consistent with the use of conditional likelihood in selective inference \cite{zollner2007overcoming,lee2016exact,heller2019post}.
 For estimation purposes, we derive for each of the interpretations the corresponding MML estimator and discuss the relations between the estimators. For performance analysis, we develop a novel MCRB, the post-model-selection MCRB (PS-MCRB), that considers both the model-selection procedure and the model misspecification. We derived the PS-MCRB for each of the interpretations and discussed their properties. Finally,  we demonstrate the proposed bounds and estimators via
 simulations of estimation after channel detection in a Gaussian linear model in a scenario where the source of the signal, whether from a known channel or not, is undetermined. 
  We show that the MML estimators under the normalized and selective inference interpretations are better than the common,  MML estimator under the naive interpretation, in terms of MSE. In addition, we show the applicability of the different versions of the PS-MCRB as appropriate bounds for these estimators that outperform the commonly used oracle CRB, which may not be a valid bound.

	The remainder of this paper is organized as follows.
	In \cref{background_Misspecification}, we present the background on estimation under model misspecification.
	In \cref{PMS_est}
 the mathematical model for estimation after model selection is presented, as well as the introduction of three interpretations to address the problem of post-model-selection estimation as an estimation under model misspecification problem.
In \cref{estimators}, we derive the MML estimators associated with the interpretations from \cref{PMS_est} as post-selection estimators and discuss their properties. 
In
\cref{performance_analysis}, we present the pseudo-true parameter vectors and the unbiasedness definition. Then, in \cref{bounds}, we present
 the PS-MCRB for the  three interpretations.
 In \cref{simulations}, we demonstrate the estimators and bounds via simulations. Finally, our conclusions appear in \cref{conclusions}.

	In the rest of this paper, vectors are denoted by boldface lowercase letters and matrices by boldface uppercase letters.
	The indicator function of an event $A$ is denoted by $\mathbbm{1}_A$ and
	the identity matrix is denoted by $\Imat$.
	The operators  $(\cdot)^T$ and  $(\cdot)^{-1}$ denote the transpose and inverse, respectively.
	The notation
 $\Amat\succeq\Bmat$  implies that $\Amat-\Bmat$ is a positive 
 semidefinite matrix.
	The $m$th element of the gradient vector  $\grad{\thvec} c$ is given by $\frac{\partial  c}{\partial \theta_m}$,
	where $\thvec=[\theta_1,\ldots,\theta_M]^{\mbox{\tiny $T$}}$, $c$ is a scalar function of $\thetavec$,
	${\grad{\thvec}}^T  c \triangleq {(\grad{\thvec}  c)}^T$, and
	$\grad{\thvec}^2  c \triangleq \grad{\thvec} {\grad{\thvec}}^T  c$.
	The notations ${\rm{E}}_{p} [\cdot]$ and ${\rm{E}}_{p} [\cdot|A]$ represent the expectation and conditional expectation w.r.t.
 given event $A$,  respectively. 

	\section{Background}
 \label{background_Misspecification}
		In this section, we briefly present background on estimation under model misspecification for the general case \cite{white1982maximum,vuong1986cramer,richmond2015parameter,fortunati2016constrained,fortunati2017performance}. 
		Classical estimation theory is based on the implicit assumption that the observation model is correctly specified \cite{kay1993fundamentals}.
		However, in practice, this assumption may be incorrect.
		Estimation under model misspecification refers to situations where the considered model, named the {\em assumed} model, is different from the  {\em{true}} model.
		model misspecification may be caused by faults or due to model relaxations that aim to reduce the estimation complexity.
  The analysis of estimation under misspecified models should take into account the statistics of both the true model and the assumed model.
		
	Non-Bayesian estimation under the misspecified setting can be formulated as follows. Let $\xvec \in \Omega_{\xvec} \subseteq \mathbb{R}^N$ be a random observation vector, which is distributed according to the pdf, $ p(\xvec)$, which represents the {\em true} observation model. Under the misspecified model, it is assumed that $\xvec $ is distributed according to the pdf $ f(\xvec; \thvec)$ that is parameterized by a deterministic parameter vector $\thvec \in \Omega_{\thvec}$. In the following,  $f(\xvec; \thvec)$ is considered to be the {\em assumed} pdf. 	
In conventional non-Bayesian estimation problems, the definition of the estimation error is straightforward: the difference between the estimator and the value of the true parameter.
		However, in estimation under misspecification, the estimation error definition is not trivial since the assumed pdf parameters, $\thvec$, do not necessarily appear in the true pdf, $p(\xvec)$. 
		To address this problem, the {\em pseudo-true} parameters are defined (see e.g. \cite{vuong1986cramer,white1982maximum,fortunati2017performance}).
			\begin{definition}(Pseudo-true parameters)	\label{Pseudo_true}		
				For a true pdf $p(\xvec)$ and an assumed pdf $f(\xvec;\thvec)$ with an assumed parameter vector $\thvec \in \Omega_{\thvec}$, the pseudo-true parameter vector is defined as
				\be \label{pseudo_true_parameter}
				{\thvecvar} \triangleq  \arg\min_{\thvec \in \Omega_{\thvec}} \kld{p(\xvec)}{f(\xvec;\thvec)},
				\ee
				where $\kld{\cdot}{\cdot}$ denotes the Kullback-Leibler divergence (KLD), which is given for general pdfs $g_1(\xvec)$ and $g_2(\xvec)$ by
				\be
    \label{kld_def}
				\kld{g_1(\xvec)}{g_2(\xvec)}\triangleq  \mathrm{E}_{g_1}\left[  \log \left(\frac{g_1(\xvec)}{g_2(\xvec)}\right) \right].
				\ee
			\end{definition}
\hspace{-0.65cm} Since $\mathrm{E}_{p}\left[  \log p(\xvec) \right]$ is not a function of $\thvec$, \eqref{pseudo_true_parameter} can be written as
		\be \label{pseudo_true_parameter2}
		{\thvecvar} =  \arg\max_{\thvec \in \Omega} \mathrm{E}_{p}\left[  \log f(\xvec;\thvec) \right].
		\ee

	It should be noted that \cref{Pseudo_true} is well defined only under the assumption that the minimization problem in \eqref{pseudo_true_parameter} has a unique minimum. 
	According to \cref{Pseudo_true}, the pseudo-true parameter vector, $\thvecvar$, is the point in 
 $\Omega_{\thvec}$ where the assumed pdf, $f(\xvec;\thvecvar)$, is the closest (in terms of KLD) to the true pdf, $p(\xvec)$.

By using \cref{Pseudo_true}, the
 misspecified MSE (MSMSE) matrix can be defined as follows:
 \be \label{MSMSE}
{\bf{MSMSE}}(\hat{ \thvec},\thvecvar)\triangleq  \mathrm{E}_p[(\hat{ \thvec}-\thvecvar)(\hat{ \thvec}-\thvecvar)^T].
 \ee
	 The misspecified-unbiasedness (MS-unbiasedness), i.e. unbiasedness under the misspecified setting, is defined as follows:
		\begin{definition}(MS-unbiasedness) \label{MS_unbiasedness_def}
			An estimator $\hat{ \thvec}$ is MS-unbiased if
			$\mathrm{E}_p[\hat{ \thvec}]=\thvecvar$,
			where $\thvecvar$ is the pseudo-true parameter vector defined in \eqref{pseudo_true_parameter}.
		\end{definition} 

	The MML estimator, also known as the quasi-ML estimator, is defined as the ML estimator of the unknown parameters under the assumed model \cite{white1982maximum,richmond2015parameter,fortunati2016misspecified,fortunati2017performance} as follows:
	\be \label{MML_conventional}
		\mml{\thvec} \triangleq \arg\max_{\thvec \in \Omega_{\thvec}} f(\xvec;\thvec).
	\ee
	Under mild regularity conditions, the MML estimator is a consistent estimator of the pseudo-true parameter, $\thvecvar$ \cite{huber1967under,white1982maximum}. 

	Finally, under mild regularity conditions,
	the MCRB is a lower bound on the MSMSE of any MS-unbiased estimator, which is given by \cite{richmond2015parameter,fortunati2016misspecified,fortunati2017performance}
	\be \label{conventional_MCRB}
{\bf{MSMSE}}(\hat{ \thvec},\thvecvar)	\succeq \Amat^{-1}(\thvecvar)\Bmat(\thvecvar)\Amat^{-1}(\thvecvar),
	\ee
	where the MSMSE is defined in \eqref{MSMSE}, 
	\be
 \label{A_def}
	\Amat(\thvec)\triangleq 	\mathrm{E}_p\left[ \grad{\thvec}^2  \log f(\xvec;\thvec)   \right]
	\ee
	is assumed to be a non-singular matrix, and
	\be
 \label{B_def}
	\Bmat(\thvec)\define \mathrm{E}_p\left[ \grad{\thvec}\log f(\xvec;\thvec) \grad{\thvec}^T\log f(\xvec;\thvec)  \right].
	\ee
	 The matrices $\Amat(\thvec)$ and $\Bmat(\thvec)$ are the Hessian form and the outer-product form Fisher information matrices (FIMs), 
  respectively.
  It should be emphasized that the likelihood functions used in the expectations in \cref{A_def,B_def} are functions of the assumed pdf, $f(\xvec;\thvec)$, while the expectation operator is computed w.r.t. the true pdf, $p(\xvec)$. Therefore, 
$	 \Amat(\thvec)\neq -\Bmat(\thvec)$ \cite{white1982maximum}.
Under a correctly-specified model, 
 where 
 $  p(\xvec)=f(\xvec; {\thvec})$, $\forall \xvec \in \Omega_{\xvec}$, ${\thvec} \in \Omega_{\thvec}$,
 the MSMSE coincides with the MSE, and $\Amat({\thvec})=-\Bmat({\thvec})$ are the identical
	 forms of the FIM. Thus, the MCRB  generalizes the  CRB for the misspecified setting.

	\section{Post-Model-Selection Estimation}
 \label{PMS_est}
 In this section, we present the problem of non-Bayesian post-model-selection estimation. In \cref{model}, the model is presented. Then, in \cref{relation_misspecified}, we present the interpretations to address this problem 
 as an estimation under model misspecification problem. 
	\subsection{Model: Estimation After Model Selection}
	\label{model}
	Let $\xvec \in \Omega_\xvec \subseteq \mathbb{R}^N$ be an observation vector, where $\Omega_\xvec$ is the observation space, with the true pdf,  $p(\xvec;\bvarphi)$, which is parameterized by the 
 deterministic parameter vector, $\bvarphi \in \Omega_{\varphi}$, where $\Omega_{\varphi}$ denotes the true parameter space. In practice, 
	the exact true model is unknown, but it is assumed that its pdf belongs to a set of candidate pdfs, $\{ f_k(\xvec;\thvec^{(k)})\}_{k=1}^K$. The $k$th pdf is parameterized by a deterministic parameter vector, $\thvec^{(k)} \in \Omega_k$, where $\Omega_{k}$ denotes the parameter space of the $k$th pdf. The $K$ different pdfs represent the $K$ different models. Finally, the vector $\thvec\triangleq [(\thvec^{(1)})^T,\ldots,(\thvec^{(K)})^T]^T \in \Theta$ is the augmented vector that contains the parameters of all candidate models, 
 where $\Theta\triangleq \Omega_1\times\ldots\times\Omega_K$.
	
	Post-model-selection estimation arises in many signal processing problems and can be described as a two-stage approach: in the first stage, the model is selected from the candidate models based on the observations. In the second stage, the parameters of the selected model are estimated based on the same observations.
	 The selection stage is conducted according to a predetermined data-based selection rule, $\Psi: \Omega_\xvec \rightarrow \{1,\ldots, K\} $, such as the MDL and AIC rules \cite{stoica2004model}.

	 We denote the deterministic sets that are associated with the selection of each model $k$  by
	\be \label{partition}
	\Am{k}\triangleq \{\xvec \in {\Omega_\xvec}: \Psi=k \},~~~
k=1,\ldots,K,
	\ee
	and assume that  $\{\Am{k}\}_{k=1}^K$ creates a disjoint partition of ${\Omega_\xvec}$, i.e. $\Am{k}\cap  \Am{m}=\emptyset$, $m\neq k$, and $\cup_{k=1}^K\Am{k}= {\Omega_\xvec}$.

	 The probability of selecting the $k$th model is denoted by
	 \be \label{p_k}
	 	p_k(\bvarphi) \triangleq \int_{\Am{k}}  p(\xvec;\bvarphi) \mathrm{d}\xvec, ~~~k=1,\ldots,K.
	 \ee 
		 In addition, we define
	 \be \label{pi_k}
	 \pi_k(\thvec^{(k)}) \triangleq \int_{\Am{k}} f_k(\xvec;\thvec^{(k)}) \mathrm{d}\xvec, ~~~k=1,\ldots,K.
	 \ee
	  It should be noted that 
   since $p(\xvec;\bvarphi)$ is unknown, the probabilities $p_k(\bvarphi)$, $k=1,\ldots,K$ are unknown. 
   For the sake of simplicity of notation, in the following, we replace $p_k(\bvarphi)$ with $p_k$.
	It should be noted that the probabilities in \eqref{p_k}, $\{p_k\}_{k=1}^K$,  are computed using  the same probability measure, and, thus,
	 \be 
	 	\sum_{k=1}^K p_k=\sum_{k=1}^K \int_{\Am{k}}  p(\xvec;\bvarphi) \mathrm{d}\xvec=\int_{\Omega_{\xvec}}  p(\xvec;\bvarphi) \mathrm{d}\xvec=1.
	 \ee
	  In contrast, 
    the probabilities 
	 $\{ \pi_k(\thvec^{(k)})\}_{k=1}^K$ are computed by integration w.r.t. a different probability measure for each $k$. Thus, 
   in the general case, the sum of the probabilities
	  \be \label{sum_pi_k}
	  	\sum\nolimits_{k=1}^K \pi_k(\thvec^{(k)})=\sum\nolimits_{k=1}^K \int_{\Am{k}} f_k(\xvec;\thvec^{(k)}) \mathrm{d}\xvec \neq 1.
	  	\ee
	  	
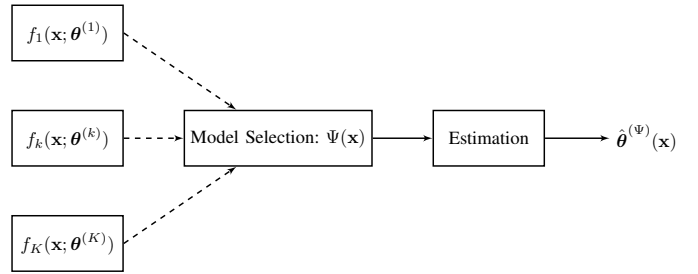
\begin{figure}[htb]   \centering \scalebox{0.7}{
	 		\begin{tikzpicture}[auto,thick, node distance=2cm,>=latex']
	 			

	 			\node [block] (fk) {$f_k(\xvec;\thvec^{(k)})$ };
	 			\node [block,above of= fk] (f1) {$f_1(\xvec;\thvec^{(1)})$ };
	 			\node [block, below of= fk] (fK) {$f_K(\xvec;\thvec^{(K)})$};
	 			\node [block,right of=fk,node distance=4cm] (q){Model Selection: {$\Psi(\xvec)$} };
		 		\draw [draw,dashed,->] (f1.east) -- (q) {};
		 		\draw [draw,dashed,->] (fk.east) -- (q) {};
		 		\draw [draw,dashed,->] (fK.east) -- (q) {};

	 			\node [block, right of=q, node distance=4cm] (Estimator) {Estimation};
	 			\node [output, right of=Estimator,node distance=3cm] (output) {$\hat{\thvec}^{(\Psi)}(\xvec)
	 				$};
	 			\draw [->] (q) -- node[name=u] {} (Estimator);
	 			
	 			\coordinate [below of=Estimator,node distance=1.2cm] (measurements) {};
	 			\draw [->] (Estimator) -- node [name=thetahat] {}(output);
	 	\end{tikzpicture}}
	 	\caption{Post-model-selection estimation scheme: first, a model is selected from a pool of candidate models based on a predetermined selection rule. Second, the unknown parameters of the selected model are estimated. Both 
   stages are performed based on the same observation vector, $\xvec$.} \label{model_scheme} 
	 \end{figure}

	  		In this paper, it is assumed that the model selection rule, $\Psi$, is predetermined, and the goal is to analyze the consequent estimation.
	  	The parameters that are estimated may be different under the different selected models.
	 Thus, 	$\hat{\thvec}^{(k)}$ denotes an estimator of the parameter vector of the $k$th model, ${\thvec}^{(k)}$,  $\forall k \in \{1,\ldots,K\}$.
	  	Finally,   	the considered post-model-selection estimation architecture is presented schematically in \cref{model_scheme}.

	 \subsection{Post-Model-Selection Estimation as Estimation Under model misspecification} \label{relation_misspecified}
	 	The problem of post-model-selection estimation described in \cref{model} has been widely discussed in the literature on selective inference (see e.g. in \cite{shen2004inference,berk2013valid,efron2014estimation,leeb2005model,lee2016exact,fithian2014optimal,belloni2013least,chaumette2005influence,chaumette2007cramer,meir2021cramer,harel2021bayesian}).
	 In this paper, we take the approach of treating it as estimation under model misspecification, which is described in \cref{background_Misspecification}.  Treating and analyzing the post-model-selection estimation as an estimation under model misspecification is far from being straightforward, since there is no clear definition of the {\em{assumed model}} in this case. This is because there are several candidate models, and the selected model differs for different observation vectors $\xvec$. 
	 	In \cref{Naive_Interpretation,Normalized_Interpretation,post_selection_Interpretation}, we describe three interpretations of post-model-selection estimation as an estimation under model misspecification
   by describing their associated {\em assumed pdf} under the misspecified model, denoted by $f_I(\xvec;\thvec)$, $f_{II}(\xvec;\thvec)$, and $f_{III}(\xvec;\thvec)$.
	 	
 	\subsubsection{Naive Interpretation} \label{Naive_Interpretation}
 	A natural approach is to treat the selected model as the assumed one and disregard the selection procedure. In this approach, if the $k$th model has been selected in the first stage, the pdf associated with the selected model, $f_k(\xvec;\thvec^{(k)})$, is considered to be the assumed pdf. Mathematically,  this implies that the assumed pdf is 
 	\be \label{f_I}
 		f_I(\xvec;\thvec)		=  f_k(\xvec;\thvec^{(k)}), ~~\forall \xvec\in\Am{k}, ~~ k=1,\ldots, K.
 	\ee 
 	By using the indicator function, \eqref{f_I} can be rewritten as
 	\be \label{f_Ia} \begin{aligned}
 		f_I(\xvec;\thvec)=  \sum_{k=1}^{K}  f_k(\xvec;\thvec^{(k)}) \ind{\xvec\in\Am{k}},~~ \forall \xvec \in \Omega_{\xvec}.
 	\end{aligned}
  	\ee 	
 	However, the function $f_I(\xvec;\thvec)$ is not a valid pdf in the general case, as it does not integrate to unity. This can be shown by integrating  $f_I(\xvec;\thvec)$ w.r.t. $\xvec$ over the observation space, $\Omega_{\xvec}$:\\
 	\be \label{not_pdf} \begin{aligned}[b]
 		 		\int_{\Omega_{\xvec}} 	f_I(&\xvec;\thvec) \mathrm{d}\xvec=	\int_{\Omega_{\xvec}} \sum_{k=1}^{K}  f_k(\xvec;\thvec^{(k)}) \ind{\xvec\in\Am{k}} \mathrm{d}\xvec\\&=
 		 		 \sum_{k=1}^K \int_{\Am{k}} 	f_k(\xvec;\thvec^{(k)}) \mathrm{d}\xvec=
 		 		 \sum_{k=1}^K \pi_k(\thvec^{(k)}),
 	\end{aligned}
 	\ee
 	where the first and second equalities are obtained by substituting \eqref{f_Ia} and by
  changing the order of summing and integration, respectively, and the last equality is obtained by substituting  \eqref{pi_k}. 
  As shown in \eqref{sum_pi_k},
  the r.h.s. of \eqref{not_pdf} is not necessarily $1$.
 	 Although  $f_I(\xvec;\thvec)$ is not a valid pdf, in practice, it can be used for the estimation approach, e.g. by using the ML estimator 
   under the selected model \cite{bozdogan1987model,nishii1989maximum,burnham1998practical,candes2007dantzig,berk2013valid,belloni2013least,ding2018model} (see more in \cref{MSL_estimator_section}). 
 	 However, in terms of analysis,
 	  referring to  $f_I(\xvec;\thvec)$ as the assumed pdf that defines the misspecified model is inappropriate, and there is no guarantee that results from \cref{background_Misspecification} hold in this case. In particular, the MCRB may not be a valid bound under this setting.

 	 	\subsubsection{Normalized Interpretation} \label{Normalized_Interpretation}
			Since the naive interpretation in \cref{Naive_Interpretation} led to a non-valid pdf, a straightforward remedy is to take the assumed pdf to be a normalized version of $f_I(\xvec;\thvec)$ from \eqref{f_Ia} as follows:
 	\be \label{f_II} 
 		f_{II}(\xvec;\thvec)
 		=\frac{1}{\alpha(\thvec)}f_I(\xvec;\thvec).
 	\ee
 	The normalization factor in \eqref{f_II}  is 
 	\be \label{alpha} 
 	\alpha(\thvec) \triangleq \sum\nolimits_{k=1}^{K}  \pi_k(\thvec^{(k)}),
 	\ee
 	where $\pi_k(\thvec^{(k)})$ is defined in \eqref{pi_k}.
 	 By substituting \eqref{f_Ia} in \eqref{f_II} we obtain that
 	 	\be \label{f_IIa} \begin{aligned}
 	 	f_{II}(\xvec;\thvec)
 	 	=\frac{1}{\alpha(\thvec)} \sum_{k=1}^{K}  f_k(\xvec;\thvec^{(k)}) \ind{\xvec\in\Am{k}},~\forall \xvec \in \Omega_{\xvec}.
 	 \end{aligned}
 	 \ee
 	 By using \eqref{f_II} and the fact that $ \alpha(\thvec)$ is not a function of $\xvec$,  it can be verified that 
 	 \beqna \label{f_IIb} 
 	 		\int_{\Omega_{\xvec}} f_{II}(\xvec;\thvec) \mathrm{d}\xvec&=&   \frac{1}{\alpha(\thvec)}\int_{\Omega_{\xvec}} 
   f_I(\xvec;\thvec)
    \mathrm{d}\xvec \nonumber
    \\
		&=&\frac{1}{\alpha(\thvec)}\sum\nolimits_{k=1}^{K}   \pi_k(\thvec^{(k)})	 		
 	 		=1,
 	 \eeqna
   where the second equality is obtained by substituting \eqref{not_pdf}, and the last equality is obtained by substituting the definition of $\alpha(\thvec) $ from \eqref{alpha}.
 	  Thus, $f_{II}(\xvec;\thvec)$,  which has non-negative values and is integrated to unity, is indeed a valid pdf and can be used as the assumed pdf in the misspecified setting. 
 	 
 	It is important to note that the normalization factor, $\alpha(\thvec)$, is a function of the parameters of all candidate models, $\thvec^{(1)},\ldots, \thvec^{(K)}$. As a result, 
  for any realization of $\xvec$, 
  $f_{II}(\xvec;\thvec)$ from \eqref{f_II}  is a function of all the unknown parameters. Thus, this interpretation creates a coupling between the unknown parameters.
 	This is as opposed to  $f_I(\xvec;\thvec)$  from \eqref{f_I}, in which if $\xvec \in \Am{k}$, then $f_I(\xvec;\thvec)$ only depends on the $k$th parameter vector, $\thvec^{(k)}$, associated with the selected model. Therefore,  $f_{II}(\xvec;\thvec)$ allows for a valid analysis, but it complicates the estimation process since, in this case, we need to estimate all $\thvec$ for any $\xvec$.
 	Furthermore, 
  we
  usually only interested in estimating the parameter under the selected model, and the parameters of the unselected models are irrelevant. Thus, this interpretation may seem cumbersome for the considered setting of post-model-selection estimation.

 	\subsubsection{Selective Inference Interpretation} \label{post_selection_Interpretation}
In order to balance competing objectives (i.e. using a valid pdf that is also tractable and reasonable), an alternative approach is proposed here. Instead of normalizing $f_I(\xvec;\thvec)$ from \eqref{f_I} using a single normalization factor, such as 
 	$\alpha(\thvec)$ from \eqref{alpha}, this approach 
  uses a separated normalization for each selected model (each marginal pdf $f_k(\xvec;\thvec^{(k)})$ on the r.h.s. of \eqref{f_Ia}). By doing so, the normalization process does not induce coupling between the parameters of the different models. Specifically, in this framework, the assumed pdf is defined as follows:
	\be \label{f_III}
		f_{III}(\xvec;\thvec)=\sum_{k=1}^K \frac{c_k}{\pi_k(\thvec^{(k)})}  f_k(\xvec;\thvec^{(k)}) \ind{\xvec\in\Am{k}},
	\ee
	 where $\{c_k\}_{k=1}^K$ are any set of constants that satisfy:
	 \renewcommand{\theenumi}{C.\arabic{enumi}} 
  \begin{enumerate}
      \item \label[condition]{cond_ck1} $c_k\geq 0,~  k=1,\ldots,K $;
      \item \label[condition]{cond_ck2}$c_k $  is not a function of   $\thvec$ and/or $\xvec$, $k =1,\ldots,K$;
      \item \label[condition]{cond_ck3} $
	\sum_{k=1}^K c_k =1$.
   \end{enumerate}
	 	\renewcommand{\theenumi}{\arabic{enumi}}

	 By using the Bayes rule and \eqref{pi_k}, one can notice that
	 \be \label{f_k_given_psi}
        \frac{f_k(\xvec;\thvec^{(k)}) }{\pi_k(\thvec^{(k)})}=f_k(\xvec|\Psi=k;\thvec^{(k)})  ,~~~~\forall \xvec\in\Am{k},
	 \ee
	which is the conditional pdf according to the $k$th model, conditioned by the event of selection in this model, $\Psi=k$.  
	 By substituting \eqref{f_k_given_psi} in \eqref{f_III}, we obtain that 
	\be \label{f_III_2}
	f_{III}(\xvec;\thvec)=\sum_{k=1}^K c_k f_k(\xvec|\Psi=k;\thvec^{(k)})\ind{\xvec\in\Am{k}}.
	\ee
	 One can verify that
	$f_{III}(\xvec;\thvec)$ is a valid pdf, since 
	\be \begin{aligned}[b]
			\int_{\Omega_\xvec}& f_{III}(\xvec;\thvec) \mathrm{d}\xvec=\int_{\Omega_\xvec}
  \sum_{k=1}^K c_k f_k(\xvec|\Psi=k;\thvec^{(k)})\ind{\xvec\in\Am{k}} \mathrm{d}\xvec
   \\=&\sum_{k=1}^K\int_{\Am{k}} c_k f_k(\xvec|\Psi=k;\thvec^{(k)}) \mathrm{d}\xvec
			=	\sum_{k=1}^K c_k =1,
	\end{aligned}
 \label{validIII}
	\ee
 where the first equality is obtained by substituting \eqref{f_III_2}, the second equality is obtained by
  changing the order of summing and integration, and the third equality is obtained by using the conditional pdf property 
  $\int_{\Am{k}}  f_k(\xvec|\Psi=k;\thvec^{(k)})\mathrm{d}\xvec=1$,  and the fact that $c_k$  are not functions of  $\xvec$ (\cref{cond_ck2})). The last equality is obtained by substituting \cref{cond_ck3}. Therefore, \eqref{validIII} implies that
   $f_{III}(\xvec;\thvec)$ is a valid pdf as long as it is a nonnegative function, which is obtained from \cref{cond_ck1}.

In general, one can choose any  coefficients $\{c_k\}_{k=1}^K$ that satisfy \cref{cond_ck1,cond_ck2,cond_ck3}. In particular, we suggest to choose  
	\be
 \label{c_k_pk}
	c_k =p_k, ~~~\forall k\in \{1,\ldots,K\},
	\ee
 where $p_k$ is 
 the true probability of selection from \eqref{p_k}.
 It can be verified that this choice satisfies \cref{cond_ck1,cond_ck2,cond_ck3}. By substituting \eqref{c_k_pk} in \eqref{f_III}, we obtain
	\be \label{f_III_ck_p_k}
		f_{III}(\xvec;\thvec)=\sum_{k=1}^K \frac{p_k}{\pi_k(\thvec^{(k)})}  f_k(\xvec;\thvec^{(k)}) \ind{\xvec\in\Am{k}}.
	\ee
Thus, for this choice the pdf
 $f_{III}(\xvec;\thvec)$ is a weighted average of the candidate pdfs, where the weight of the marginal pdf of the $k$th model, $f_k(\xvec;\thvec^{(k)})$, is the likelihood ratio of 
 the correct and the candidate models, $\frac{p_k}{\pi_k(\thvec^{(k)})}$. While the probabilities $\{p_k\}_{k=1}^K$ are unknown,
   $f_{III}(\xvec;\thvec)$ from \eqref{f_III_ck_p_k} lead to a valid
 estimator and bound, as shown in  \cref{PSML,bound_III_subsec}.

 	It can be seen that for a given $\xvec \in \Am{k}$, similar to $f_I(\xvec;\thvec)$ in \eqref{f_Ia},  the assumed pdf in \eqref{f_III}, $f_{III}(\xvec;\thvec)$,  is only a function of the parameter vector, $\thvec^{(k)}$,  associated with the selected model.  Thus, the assumed pdfs $f_I(\xvec;\thvec)$ and $f_{III}(\xvec;\thvec)$ are {\em coherent} with the selection approach. Coherency in estimation after model selection refers to the property that the estimation approach accounts for the model selection
  (see more in \cite{meir2021cramer,harel2021bayesian,harel2021post}). In our case, this coherency means that if it is assumed that the observations obey the $k$th model, then the assumed pdf is only determined by  $\thvec^{(k)}$. This is as opposed to $f_{II}(\xvec;\thvec)$ in \eqref{f_II}, which is a function of the augmented vector, $\thvec$. In addition, 
 	 	 in previous works on selective inference, it was shown that the conditional pdf of the observation conditioned on the selection is the appropriate pdf for post-selection estimation and analysis (see e.g.\cite{fithian2014optimal,lee2016exact,heller2019post,zollner2007overcoming,routtenberg2016estimation}). Thus,  $f_{III}(\xvec;\thvec)$ corresponds with the existing theory. 
   
 	To conclude this subsection, we note that for the special case where there is a single candidate model, i.e. $K=1$ and $\Am{1}=\Omega_{\xvec}$,  all the interpretations above coincide: $f_I(\xvec;\thvec)=f_{II}(\xvec;\thvec)=f_{III}(\xvec;\thvec)$, since  \cref{pi_k,alpha} imply that in this case $\alpha(\thvec)=\pi_1(\thvec)=1$. Moreover, in this case, the considered scheme is reduced to conventional parameter estimation under model misspecification \cite{white1982maximum,richmond2015parameter,fortunati2016misspecified}, where the single candidate model is represented by the pdf $f_1(\xvec;\thvec^{(1)})$.
 	If, in addition, this only candidate is the true model, i.e. if $f_1(\xvec;\thvec^{(1)})=p(\xvec;\bvarphi)$, the considered scheme is reduced to the conventional non-Bayesian parameter estimation problem.
	 \section{Post-Model-Selection Estimators}\label{estimators}
	 In this section, we introduce post-model-selection estimators.
  We start by presenting the oracle ML estimator as a benchmark. 
  Then,
	 in \cref{MSL_estimator_section,MSNL_sec,PSML}, we present the MML estimators as post-model-selection estimators
  according to the interpretations presented above in \cref{relation_misspecified}.

	 The oracle ML estimator is the ML estimator based on the true (unknown) model, i.e.
	 \be \label{OML_est}
	 \oml{\bvarphi}\triangleq \arg\max_{\bvarphismall \in \Omega_{\varphi}} p(\xvec;\bvarphi).
	 \ee 
	 The oracle ML estimator is hypothetical, as it assumes knowledge of the true model, which is unknown in the considered setting.
  Thus, it does not take the form of a practical post-model-selection estimator from \eqref{generic_post_selection_estimator}.
	 Nevertheless, the oracle ML estimator is a common benchmark for the analysis of estimators under misspecification \cite{ben2010cramer,elad2010sparse}, and therefore is included here for the sake of completeness.

	 
	 \subsubsection{Maximum Selected Likelihood (MSL) Estimator} \label{MSL_estimator_section}
	 An intuitive post-model-selection estimator is the MSL estimator obtained by first selecting a model and then setting the estimator to be the ML estimator of the selected model. 
	Therefore,
the MSL estimator is given by
\be \label{MSL}
\msl{\thvec}^{(k)}\triangleq\arg\max_{\thvec^{(k)} \in \Omega_k}    f_I(\xvec;\thvec) ,~ \forall \xvec \in \Am{k}.	
\ee
Since by the definition in \eqref{f_I}, $f_I(\xvec;\thvec)= f_k(\xvec;\thvec^{(k)}) ,~ \forall \xvec \in \Am{k}$,
the MSL estimator can be written as
\be \label{MSL_2}
\msl{\thvec}^{(k)}=\arg\max_{\thvec^{(k)} \in \Omega_k}    f_k(\xvec;\thvec^{(k)}) ,~ \forall \xvec \in \Am{k}.	
\ee
Therefore, one may consider the MSL estimator as the MML estimator under the naive interpretation presented in \cref{Naive_Interpretation}. 
Although $f_I(\xvec;\thvec)$ is not a valid likelihood function (as explained in \cref{Naive_Interpretation}), the MSL estimator is well defined and widely used in practice {\cite{bozdogan1987model,nishii1989maximum,burnham1998practical,candes2007dantzig,berk2013valid,belloni2013least,ding2018model}}.

	 \subsubsection{Maximum Selected Normalized Likelihood (MSNL) Estimator}\label{MSNL_sec}

	 The MSNL estimator is given by the maximization of the assumed likelihood from \cref{f_II}, as follows:
 \be \label{MSNL_est}
	\msnl{\thvec} \triangleq \arg\max_{\thvec \in \Theta}    f_{II}(\xvec;\thvec),
\ee
which is the MML estimator  defined in \eqref{MML_conventional}, where \cref{f_II} is the assumed pdf.
By substituting \eqref{f_IIa} in \eqref{MSNL_est} and using the fact that 
the log function  is a monotonically increasing function,
we obtain that the MSNL estimator can be written as
\be \label{MSNL_est1}
\msnl{\thvec} = \arg\max_{\thvec \in \Theta}   \log f_k(\xvec;\thvec^{(k)})-\log \alpha(\thvec) ,~ \forall \xvec \in \Am{k}.	
\ee
 	The maximization in \eqref{MSNL_est1} is w.r.t.  $\thvec$, which also includes the parameters of the unselected models, $\thvec^{(l)}$, $l \neq k$, $\forall \xvec \in \Am{k}$.
 	This is since the normalization factor, $\alpha(\thvec)=\sum_{l=1}^K \pi_l(\thvec^{(l)})$ from \eqref{alpha}, is a function of the parameters of all candidate models. Nevertheless, 
 $\alpha(\thvec)$ is not a function of the observation vector, $\xvec$, and is determined by the selection rule, $\Psi$.
Thus, since the log function is a monotonically increasing function, the maximization of $-\log\alpha(\thvec)$ w.r.t.  $\thvec^{(l)}$, $l\neq k$, in \eqref{MSNL_est1} can be obtained by the minimization of $\pi_l(\thvec^{(l)})$. Thus, the MSNL estimation of the unselected parameters is given by
 	\be \label{pi_l_min}
 		\msnl{\thvec}^{(l)} = \arg\min_{\thvec^{(l)} \in \Omega_l} \pi_l(\thvec^{(l)}),~~~l\neq k, ~\forall \xvec \in \Am{k}.
 	\ee
  The unselected parameters can be interpreted as nuisance parameters \cite{bar2015partI},
 where we are only interested in the following minimum value (and not the estimators in   \eqref{pi_l_min}):
 	\be \label{pi_l_min1}
 	 	\underline{\pi_k} \triangleq \min_{\thvec^{(k)} \in \Omega_k} \pi_k(\thvec^{(k)}), ~~ k\in \{1,\ldots,K\} .
 	 	\ee 
Substitution of the minimal values from \cref{pi_l_min1} in \cref{alpha} results in
 	\be\label{alpha_k} 
 			\alpha_k(\thvec^{(k)})
 			\triangleq \pi_k(\thvec^{(k)}) +\sum_{l \neq k} \underline{\pi_l}.  
 	\ee
 By substituting \eqref{alpha_k} in \eqref{MSNL_est1}, we obtain 
\be \label{MSNL_est_k} 
\msnl{\thvec}^{(k)} =
\arg\max_{\thvec^{(k)} \in \Omega_k} \log f_k(\xvec;\thvec^{(k)})-\log \alpha_k(\thvec^{(k)}),~~\forall\xvec \in \Am{k}.
\ee
  
 	To conclude, if the $k$th model has been selected, the parameters of all the other models (``unselected parameters")  can be estimated via \eqref{pi_l_min}. Then, the parameters of the selected model are estimated via the estimator in \eqref{MSNL_est_k}, which can be interpreted as a penalized ML estimator, where the penalty is determined by the selection rule and the probabilities under the other candidate models. Thus, in contrast to the MSL estimator in \eqref{MSL},
  the MSNL estimator also integrates into the estimation the alternative models that can be selected.
  Finally, it is important to mention that the minimization in \cref{pi_l_min,pi_l_min1} can be performed in advance (offline), since \eqref{pi_l_min} is not a function of the observation vector, $\xvec$.

	 	 \subsubsection{Post-Selection ML (PSML) Estimator}\label{PSML}	 	 	 	 
	 	 The PSML estimator is given by maximizing the assumed likelihood from \cref{f_III}. That is, it is the MML estimator under the selective inference interpretation from \cref{post_selection_Interpretation} given by
				\be \label{PSML_est} 
					\psml{\thvec}^{(k)}\triangleq\arg\max_{\thvec^{(k)} \in \Omega_k}    f_{III}(\xvec;\thvec^{(k)}) ,~ \forall \xvec \in \Am{k}.	
				\ee
				By substituting \eqref{f_III_2} in \eqref{PSML_est}, and since $\{c_k\}_{k=1}^K$ are positive constants (see \cref{cond_ck1}), independent of $\thvec$ (see \cref{cond_ck3}),   the maximization  in \eqref{PSML_est} is equivalent to 
				\be \label{PSML_est_k}
				\psml{\thvec}^{(k)} =\arg\max_{\thvec^{(k)} \in \Omega_k} f_k(\xvec|\Psi=k;\thvec^{(k)}) ,~ \forall \xvec \in \Am{k}.
				\ee
				Since the log function is a monotonically increasing function, by using the Bayes rule in \eqref{f_k_given_psi},  \cref{PSML_est_k} can be written as
				\be \label{PSML_est_k_log}
				\psml{\thvec}^{(k)} =\arg\max_{\thvec^{(k)} \in \Omega_k} \log f_k(\xvec;\thvec^{(k)})-\log\pi_k(\thvec^{(k)}) ,~ \forall \xvec \in \Am{k}.
				\ee
                The PSML estimator is independent of the constants $\{c_k\}_{k=1}^K$. Therefore,
				similar to the MSL estimator, for any $\xvec \in \Am{k}$, $f_{III}(\xvec;\thvec)$ is only a function of the parameters of the $k$th model, $\thvec^{(k)}$ in contrast with the MSNL estimator. 
				The PSML estimator 
    can be interpreted as a penalized ML estimator \cite{goodd1971nonparametric,nitzan2018cramer}, where the penalty term is $-\log\pi_k(\thvec^{(k)})$. This penalty term compensates for the selection approach, which increases the uncertainty in the estimation. However, since we do not know the true model, this probability is only an estimate of the probability of selection under the $k$th model, as defined in \eqref{pi_k}. Since the penalty term is not a pdf w.r.t. $\thvec^{(k)}$, 
    the PSML estimator does not have a Bayesian interpretation. 
				
			  	 Estimators that are based on maximizing conditional likelihood functions were shown to be more appropriate for post-selection estimation problems \cite{routtenberg2016estimation,heller2019post}. Thus, the PSML estimator, which is the MML estimator under the selective inference interpretation, is expected to yield good MSE performance, as demonstrated numerically in \cref{simulations}. 
			  	 
			  	 	The following remarks describe some relations between the estimators and special cases where these estimators coincide.
			  	 	\begin{remark}\label{signle_candidate_estimators} 
			  	 			At the end of \cref{relation_misspecified}, it is shown that for the special case of a single candidate model, $K=1$, the three interpretations coincide, i.e. $f_I(\xvec;\thvec)=f_{II}(\xvec;\thvec)=f_{III}(\xvec;\thvec)$. Thus,  in this case, the MSL, MSNL, and PSML estimators from \cref{MSL,MSNL_est,PSML_est} are all reduced to the conventional MML estimator in \cref{MML_conventional}.
			  	 	\end{remark}
			  	  \begin{remark}\label{MSNL_PSML_coincide_MSL}
			  	 In the case where $\pi_k(\thvec^{(k)})$ is not a function of $\thvec^{(k)}$ for a given $k$, it can be seen that the maximizations in \cref{MSNL_est_k,PSML_est_k_log} are equivalent to the maximization in \eqref{MSL} for $ \xvec \in \Am{k}$. Therefore, in this case, the MSL, MSNL, and PSML estimators coincide under the selection of the $k$th model. 
			  	 	\end{remark}
			  	 \begin{remark}\label{MSNL_coincide_PSML}
			  	 	From the definition in \eqref{pi_k}, the probabilities satisfy $\pi_k(\thvec^{(k)})\in [0,1]$. Therefore, in the case where for some $k$, $\pi_l(\thvec^{(l)}) $ achieves its minimum at zero   $\forall l\neq k$, i.e. $\underline{\pi_l} =0$   $\forall l\neq k$,  then the  MSNL estimator from \eqref{MSNL_est_k} coincides with the PSML estimator  in  \eqref{PSML_est_k_log} under the selection of the $k$th model.
       This can be verified by substitution of \eqref{pi_l_min1} with $\underline{\pi_l} =0$ $\forall l\neq k$ in
			  	 	the MSNL estimator in \eqref{MSNL_est_k}, which results in
			  	 	$
			  	 	\msnl{\thvec}^{(k)} =
				\psml{\thvec}^{(k)} 
			  	 	$.
			  	 	\end{remark}

\section{Post-Model-Selection Performance Analysis} \label{performance_analysis}
	In the context of post-model-selection estimation as presented in \cref{model}, 
if the $k$th model has been selected, we focus on estimating the parameter vector $\thvec^{(k)}$ of the selected $k$th model, where the parameters from the unselected models can be interpreted as nuisance parameters. 
Hence, in order to analyze post-model-selection estimators that may estimate different quantities for different observation vectors, we introduce the $k$th-MSE, the associated pseudo-true parameter, and the post-selection unbiasedness in Subsection \ref{unbiased_subsec}. Subsequently,  we derive the pseudo-true parameter vectors for the different interpretations in \cref{pseudo_true_section}.

     \subsection{Post-model-selection MSE
     }
    \label{unbiased_subsec} 
    
    
    The $k$th-MSE for estimating ${\thvec}^{(k)}$ under the selection of the  $k$th  model is the following $|\Omega_k|\times |\Omega_k|$ matrix:
    \be \label{MSE_k__} \begin{aligned}[b]
    	{\bf{MSE}}^{(k)}(&\hat{\thvec}^{(k)},\thvec^{(k)}) \\ =& \mathrm{E}_p[  (\hat{\thvec}^{(k)}-\thvec^{(k)})(\hat{\thvec}^{(k)}-\thvec^{(k)})^T |\Psi=k ]	.
    \end{aligned}
    \ee
  Thus, for each $k$, the dimensions of the $k$th-MSE may be different.
  
  The MCRB in \eqref{conventional_MCRB} is evaluated at the pseudo-true parameter. Thus,  in order to discuss the post-model-selection performance,  
  it is crucial to define the pseudo-true parameter vectors. The following definition generalizes the pseudo-true parameter vector definition in \cref{Pseudo_true} to the misspecified case.
     \begin{definition}{($k$th pseudo-true parameter vector)} \label{k_pseudo_true}
     	Let $f(\xvec;\thvec)$ be a generally assumed pdf in the post-model-selection estimation setting. The $k$th pseudo-true parameter vector w.r.t. to $f(\xvec;\thvec)$, $\thvecvar^{(k)}$, is defined as follows:
     	\be \label{k_pseudo_true_parameter_kld}
     	\thvecvar^{(k)} \triangleq  \arg\min_{\thvec^{(k)} \in \Omega_{k}} \kld{p(\xvec|\Psi=k;\bvarphi)}{f(\xvec;\thvec)},
     	\ee
      where, according to the Bayes rule and similar to  \eqref{f_k_given_psi},
      \be \label{p_given_psi}
        p(\xvec|\Psi=k;\bvarphi) =\frac{p(\xvec;\bvarphi)}{p_k} ,~~~~\forall \xvec\in\Am{k}.
	 \ee
       \end{definition}
       In other words,  the $k$th pseudo-true parameter vector is obtained by minimization of the KLD between the true pdf conditioned by the selection $\Psi=k$, $p(\xvec;\bvarphi|\Psi=k)$, and the assumed pdf
. According to this definition, $\thvecvar^{(k)}$ is the point that is {\em the closest} to the ground truth, given the $k$th selection.
       It should be noted that the support of 
$p(\xvec|\Psi=k;\bvarphi)$ is $\Am{k}$, which is included in the support of $f(\xvec;\thvec)$, $\Omega_\xvec$, and thus, the KLD on the r.h.s. of \eqref{k_pseudo_true_parameter_kld} is well defined. In addition,
       \beqna 
       \label{kld_def2} 
             \kld{p(\xvec|\Psi=k;\bvarphi)}{f(\xvec;\thvec)}=\hspace{3.25cm}\nonumber
             \\
       \mathrm{E}_{p}\left[  \log p(\xvec|\Psi=k;\bvarphi)  |\Psi=k \right]-\mathrm{E}_{p}\left[  \log f(\xvec;\thvec) |\Psi=k \right].
       \eeqna
     This definition is equivalent to  \cref{Pseudo_true}, but with conditional pdfs, which fits the conditional MSE in \eqref{MSE_k__}.
     As in \cref{Pseudo_true}, since $p(\xvec|\Psi=k)$ is not a function of $\thvec^{(k)}$,  the minimization in \cref{k_pseudo_true_parameter_kld} is reduced to 
     	\be \label{k_pseudo_true_parameter}
     	\thvecvar^{(k)} =  \arg\max_{\thvec^{(k)} \in \Omega_{k}} \mathrm{E}_p\left[ \log f(\xvec;\thvec) |\Psi=k\right].
     	\ee

    In a similar manner to the conventional estimation under model misspecification, by using \cref{k_pseudo_true} we can define the $k$th-MSE of an estimator under the assumed pdf $f(\xvec;\thvec)$ as the $k$th-MSE from \cref{MSE_k__} evaluated at the $k$th pseudo-true parameter vector, $\thvecvar^{(k)}$,  
    from \eqref{k_pseudo_true_parameter} as follows:
    \be \label{MSE_k} \begin{aligned}[b]
    	{\bf{MSE}}^{(k)}(&\hat{\thvec}^{(k)},\thvecvar^{(k)}) \\ =& \mathrm{E}_p\left[  (\hat{\thvec}^{(k)}-\thvecvar^{(k)})(\hat{\thvec}^{(k)}-\thvecvar^{(k)})^T |\Psi=k \right]	.
    \end{aligned}
    \ee

    Based on the conditional MSE in \eqref{MSE_k},  we can define the associated post-selection MS-unbiasedness (PSMS-unbiasedness) definition,    which generalizes the MS-unbiasedness from \cref{MS_unbiasedness_def}.
    \begin{definition}{(PSMS-unbiasedness) } \label{PSMS_unbiasedness_def}
    	An estimator of the $k$th parameter vector, $\hat{ \thvec}^{(k)}$, is a PSMS-unbiased estimator
    	w.r.t. 
    	a generic assumed pdf $f(\xvec;\thvec)$ 
    	if
    	\be \label{PSMS_unbiasedness}
    	\mathrm{E}_p[\hat{ \thvec}^{(k)}|\Psi=k]=\thvecvar^{(k)},
        \vspace{-0.1cm}
    	\ee 
    	where $\thvecvar^{(k)}$ is defined in \eqref{k_pseudo_true_parameter_kld}. 
    \end{definition} 
    This unbiasedness definition is similar to the definition of selective unbiasedness ($\Psi$-unbiasedness)   from \cite{meir2021cramer,routtenberg2016estimation,harel2020low}. However, here we also incorporate the model misspecification by taking the $k$th pseudo-true parameter
    vector w.r.t. the assumed pdf, $f(\xvec;\thvec)$, into account.
    It can be seen that for the special case of a single candidate model ($K=1$), the PSMS-unbiasedness \eqref{PSMS_unbiasedness} coincides with the MS-unbiasedness under misspecification in 
    \cref{MS_unbiasedness_def}. 
    Finally, the PSMS-unbiasedness definition from \cref{PSMS_unbiasedness_def} can be interpreted as the unbiasedness in the Lehmann sense \cite{Lehmann1998} w.r.t. the $k$th-MSE defined in \eqref{MSE_k}. This can be seen by setting the $k$th-MSE from \eqref{MSE_k} in the definition of Lehmann unbiasedness.

 	\subsection{Post-Model-Selection Pseudo-True Parameter Vectors} \label{pseudo_true_section}
 
 	In \cref{k_pseudo_true}, the $k$th pseudo-true parameter vector is defined for a generic assumed pdf, $f(\xvec;\thvec)$. 
 	In this subsection, we derive the pseudo-true parameter vectors of the three interpretations of post-model-selection estimation.
	 	 		\subsubsection{Naive Interpretation}		
	 	By substituting $f(\xvec;\thvec)=f_I(\xvec;\thvec)$ from \eqref{f_Ia} in  \cref{k_pseudo_true_parameter}, we obtain that the $k$th pseudo-true parameter vector under the naive interpretation is given by
	 	\be \label{k_pt_I}
      \vspace{-0.1cm}
	 	\thvecvar^{(k)}_I =  \arg\max_{\thvec^{(k)} \in \Omega_{k}} \mathrm{E}_p\left[ \log f_k(\xvec;\thvec) |\Psi=k\right],
   \vspace{-0.1cm}
	 	\ee
   $k =1,\ldots,K$. 
      As explained in \cref{Naive_Interpretation}, $f_I(\xvec;\thvec)$ is not a valid pdf. 
      Thus, using \cref{Pseudo_true} will not result in a KLD measure in the conventional sense. 
      On the contrary, \cref{k_pseudo_true} implies that $\thvecvar^{(k)}_I$ is  the minimizer of the valid KLD between $f_k(\xvec;\thvec^{(k)})$ and the true {\em conditional} pdf, $p(\xvec;\bvarphi|\Psi=k)$ in the support $\Am{k}$. 
 The relationship between the two definitions of pseudo-true parameter vectors for the naive interpretation is described in the following claim.
\begin{claim}
\label{claim1}
    Let $	{\thvecvar}_{I}\triangleq [({\thvecvar}^{(1)}_{I})^T,\ldots,({\thvecvar}^{(K)}_{I})^T]^T $, 
    where ${\thvecvar}^{(k)}_{I}$ is the $k$th pseudo-true parameter vector under the naive interpretation according to \cref{k_pseudo_true}. Then, under the naive interpretation, ${\thvecvar}_{I}$ is the pseudo-true parameter vector according to 
\cref{Pseudo_true}.
\end{claim}
\begin{IEEEproof}
    By substituting $f(\xvec;\thvec)=f_I(\xvec;\thvec)$ in \eqref{pseudo_true_parameter2}, 
	 		we obtain that,
    according to  \cref{Pseudo_true}, 
    the pseudo-true parameter under the first, naive  interpretation from \cref{Naive_Interpretation} is
	 		\be \label{I_pseudo_parameter}
	 		\tilde{\thvecvar}_{I} \triangleq  \arg\max_{\thvec \in \Theta} \mathrm{E}_{p}\left[  \log f_I(\xvec;\thvec) \right].
	 		\ee
	 	By substituting \eqref{f_Ia} in   the r.h.s of \eqref{I_pseudo_parameter}, we obtain that 
	 		\be
    \label{p_I_KL}\begin{aligned}[b]
	 			\mathrm{E}_p&\left[ \log f_I(\xvec;\thvec) \right]\\&=\int_{\Omega_{\xvec}}p(\xvec;\bvarphi)\log\left(\ \sum\limits_{k=1}^{K}  f_k(\xvec;\thvec^{(k)}) \ind{\xvec\in\Am{k}}   \right) \mathrm{d}\xvec\\
	 			&=\sum\limits_{k=1}^{K}\int_{\Am{k}}p(\xvec;\bvarphi) \log f_k(\xvec;\thvec^{(k)})  \mathrm{d}\xvec,
	 		\end{aligned}
       \vspace{-0.1cm}
	 		\ee
 where the last equality is obtained by
  changing the order of summing and integration, and using  $\Am{k}\cap  \Am{m}=\emptyset$,  $m\neq k$, and the indicator function properties.
 By substituting 
 \eqref{p_given_psi} in \eqref{p_I_KL} and using the conditional expectation definition, one obtains
\be
    \label{p_I_KL2}
    \begin{aligned}[b]
	 			\mathrm{E}_p\left[ \log f_I(\xvec;\thvec) \right]=\sum\limits_{k=1}^{K}p_k\mathrm{E}_p[ \log f_k(\xvec;\thvec^{(k)})|\Psi=k ].
	 		\end{aligned}
	 		\ee
By substituting \eqref{p_I_KL2} in  \eqref{I_pseudo_parameter}, we obtain that the maximization in  \eqref{I_pseudo_parameter}  is separable w.r.t. the parameters of the different models. Thus,  
it can be implemented by solving the  $K$ maximization problems in \eqref{k_pt_I}, i.e.
	$\tilde{\thvecvar}_{I}={\thvecvar}_{I}$, which completes the proof.
\end{IEEEproof}

	 		\subsubsection{Normalized Interpretation}
		By substituting $f(\xvec;\thvec)=f_{II}(\xvec;\thvec)$ from \eqref{f_IIa} in  \cref{k_pseudo_true_parameter}, we obtain that the $k$th pseudo-true parameter vector under the normalized interpretation is 
	\be \label{k_pt_II}
	\thvecvar^{(k)}_{II} =  \arg\max_{\thvec^{(k)} \in \Omega_{k}} \mathrm{E}_p\left[ \log f_k(\xvec;\thvec) |\Psi=k\right]- \log \alpha (\thvec).
	\ee
	In a similar manner to the MSNL estimator in \cref{MSNL_est1}, $\log \alpha(\thvec)$ is a function of all the parameters $\thvec^{(1)},\ldots,\thvec^{(K)}$. Nevertheless, since $\thvec^{(l)}$, $l\neq k$, appears only in 
	$\alpha(\thvec)$, and since
		 the logarithmic function is a monotonically increasing function, the maximization is obtained by substituting the minimal values of $\pi_l$ from \cref{pi_l_min1}
	 . Therefore, \cref{k_pt_II} is equivalent to 
	\be \label{k_pt_II_final}
	\thvecvar^{(k)}_{II} \hspace{-0.1cm}=  \hspace{-0.1cm}\arg\hspace{-0.1cm}\max_{\thvec^{(k)} \in \Omega_{k}} \hspace{-0.15cm}\mathrm{E}_p\left[ \log f_k(\xvec;\thvec) |\Psi=k\right]- \log \alpha_k (\thvec^{(k)}),
	\ee
 where $\alpha_k (\thvec^{(k)})$ is defined in \cref{alpha_k}.
  The maximization in \cref{k_pt_II_final} is similar to  \cref{k_pt_I} with an additional penalty term, $\log \alpha_k (\thvec^{(k)})$.
	 		\subsubsection{Selective Inference Interpretation}
	 	By substituting 
   $f(\xvec;\thvec)=f_{III}(\xvec;\thvec)$
  from  \eqref{f_III_2} in  \cref{k_pseudo_true_parameter}, we obtain that the $k$th pseudo-true parameter vector under the selective inference interpretation is given by 
	 	\be \label{k_pt_III}
   \vspace{-0.1cm}
	 	\thvecvar^{(k)}_{III} \hspace{-0.1cm}= \hspace{-0.1cm} \arg\hspace{-0.1cm}\max_{\thvec^{(k)} \in \Omega_{k}} \hspace{-0.15cm}\mathrm{E}_p\left[ \log f_k(\xvec;\thvec) |\Psi=k\right]- \log \pi_k (\thvec^{(k)}).
   \vspace{-0.1cm}
	 	\ee
The following claim describes the relationship between this definition and
the pseudo-true parameter vector in \cref{Pseudo_true}.
\begin{claim}
\label{claim2}
    Let ${\thvecvar}_{III}\triangleq [({\thvecvar}^{(1)}_{III})^T,\ldots,({\thvecvar}^{(K)}_{III})^T]^T $, 
    where ${\thvecvar}^{(k)}_{III}$ is the pseudo-true parameter vector of the selective inference interpretation according to \cref{k_pseudo_true} under the $k$th selected model. 
    Then, under the selective inference interpretation, ${\thvecvar}_{III}$ is also the pseudo-true parameter vector  according to 
\cref{Pseudo_true}.
\end{claim}
\begin{IEEEproof}
By substituting $f(\xvec;\thvec)=f_{III}(\xvec;\thvec)$ in \eqref{pseudo_true_parameter2},  we obtain that,   according to  \cref{Pseudo_true},   the pseudo-true parameter under the selective inference interpretation 
is
	 		\be \label{III_pseudo_parameter}
	 		\tilde{\thvecvar}_{III} \triangleq  \arg\max_{\thvec \in \Theta } \mathrm{E}_{p}\left[  \log f_{III}(\xvec;\thvec) \right].
	 		\ee
	 	 By substituting 	  \cref{f_III} in \cref{III_pseudo_parameter} we obtain that
	 		\beqna
    \label{p_III_KL2}
	 			\mathrm{E}_p\left[ \log f_{III}(\xvec;\thvec) \right]\hspace{5.25cm}\nonumber\\
	 			=\int_{\Omega_{\xvec}}\hspace{-0.2cm}p(\xvec;\bvarphi)\log\left(\hspace{-0.1cm}\ \sum\limits_{k=1}^{K} c_k f_k(\xvec|\Psi=k;\thvec^{(k)}) \ind{\xvec\in\Am{k}}   \right) \mathrm{d}\xvec\nonumber\\
	 			=\sum\limits_{k=1}^{K}\int\limits_{\Am{k}}p(\xvec;\bvarphi) \log f_k(\xvec|\Psi=k;\thvec^{(k)})  \mathrm{d}\xvec+p_k\log c_k\nonumber\\
	 			=\sum\limits_{k=1}^{K}p_k\left(\mathrm{E}_p[ \log f_k(\xvec|\Psi=k;\thvec^{(k)})|\Psi=k ]+\log c_k \right),
	 		\eeqna
  where the second equality is obtained by
  changing the order of summing and using the property of summing over non-overlapping events, as in  \eqref{p_I_KL}. The last equality is obtained 
 by substituting the conditional pdf from \eqref{p_given_psi} in \eqref{p_I_KL} and using the conditional expectation definition.   
 By substituting \eqref{p_III_KL2} in  \eqref{III_pseudo_parameter}, we obtain that the maximization in  \eqref{III_pseudo_parameter}  is separable w.r.t. the parameters of the different models, and thus  
it can be implemented by solving the $K$ maximization problems in \eqref{k_pt_III}, i.e.
	$\tilde{\thvecvar}_{III}={\thvecvar}_{III}$, which completes the proof.
\end{IEEEproof}
    

 \section{Post-Model-Selection Misspecified Cram$\acute{\text{e}}$r-Rao-Type Lower Bounds}\label{bounds}
 
The MCRB presented in \cref{background_Misspecification} is a Cram$\acute{\text{e}}$r-Rao-type bound on the MSMSE that takes into account model misspecification.
	 	In this section,
   we derive the MCRBs that incorporate the model selection 
   as the form of model misspecification.
    
    Based on the regularity conditions of the MCRB for the general case (see e.g. \cite{vuong1986cramer,fortunati2017performance}), 
	 	we define the following regularity conditions 
   for the post-model-selection scheme with a general assumed pdf, $f(\xvec;\thvec)$, which can be e.g. one of the 
      assumed pdfs, $f_i(\xvec;\thvec)$, $i=I,II,III$. 
	 	\renewcommand{\theenumi}{C.\arabic{enumi}} 
    \renewcommand{\theenumi}{RC.\arabic{enumi}} 
	 	\begin{enumerate}[wide, labelwidth=!, labelindent=!]
	 		\item \label[condition]{cond1}
	 		The maximum of $\mathrm{E}_p[\log f(\xvec;\thvec)|\Psi=k]$ w.r.t. $\thvec^{(k)}$ from 
    \eqref{k_pseudo_true_parameter}
    is unique, $\forall k\in\{1,\ldots,K\}$.
	 		\item \label[condition]{cond2} 
	 		The log-likelihood function, $\log f(\xvec;\thvec)$, is twice differentiable function,
  and the functions $ \left|\der{\log f(\xvec;\thvec)}{\theta_i}\right|$ and
$\left|\der{^2\log f(\xvec;\thvec)}{\theta_i^2}\right|$, $i=1,\ldots, |\Omega_{\Theta}|$, are dominated by a function 
$m(\xvec)$, which is a  square-integrable function w.r.t. the true conditional pdf, $p(\xvec|\Psi=k;\bvarphi)$ from \eqref{p_given_psi}, $\forall k\in\{1,\ldots,K\}$.

	 		\item \label[condition]{cond4}
	 		There is a  neighborhood of a general $k$th pseudo-true parameter vector, ${\thvecvar}^{(k)}$, such that
	 		\be
	 		\left( \frac{1}{f(\xvec;\thvec)}\left|\der{\log f(\xvec;\thvec)}{\theta^{(k)}_i}\right|\right)\eval{\thvec^{(k)}={\thvecvar}^{(k)}} \leq m(\xvec),
	 		\ee
	 		where $m(\xvec)$ is square-integrable function w.r.t. $p(\xvec|\Psi=k;\bvarphi)$, defined in \eqref{p_given_psi}.
	 		
	 		\item \label[condition]{cond3}
	 		The $k$th post-model-selection Hessian form information matrix, evaluated at the $k$th pseudo-true parameter vector
	 		\be \label{Amat}
	 		\Amat^{(k)}(\thvecvar^{(k)})\triangleq 
	 		\left. \mathrm{E}_p\left[ \grad{\thvec^{(k)}} ^2 \log f(\xvec;\thvec) | \Psi=k  \right]
	 		\right|_{\thvec^{(k)}=\thvecvar^{(k)}},
	 		\ee 
	 		is a $|\Omega_k| \times |\Omega_k|$ non-singular matrix, where 
	 		the conditional expectation in \eqref{Amat} is obtained by integration w.r.t. the pdf $p(\xvec|\Psi=k;\bvarphi)$ from \eqref{p_given_psi}, $\forall k\in\{1,\ldots,K\}$.	 	
	 	\end{enumerate}
\cref{cond1} ensures the uniqueness of the pseudo-true parameter vector, ${\thvecvar}^{(k)}$.
\cref{cond2,cond4} enable differentiation under the integral sign of the conditional expectation of any finite-variance function of $\xvec$.
\cref{cond3} ensures that the inverse of $\Amat^{(k)}(\thvecvar^{(k)})$ from \cref{Amat} is well defined.
 All regularity conditions are w.r.t. the conditional pdf
in \eqref{p_given_psi}, which accounts for the selection approach. 
   

 
 In the following, we present the PS-MCRBs under the three interpretations
		by using the definition of the MCRB in \eqref{conventional_MCRB} and substituting the assumed pdfs from \cref{relation_misspecified}, and their pseudo-true parameter vectors from \cref{pseudo_true_section}.
		\begin{Theorem}{\label{PS_MCRB_k}}($k$th PS-MCRB)
			Let $f(\xvec;\thvec)$ be a general  assumed pdf for a post-model-selection scheme with a selection rule $\Psi$ that satisfy  \cref{cond1,cond2,cond3,cond4}. The $k$th-MSE of any finite variance, PSMS-unbiased estimator, $\hat{\thvec}^{(k)}$, satisfies 
			\be
   \label{bound1}
			{\bf{MSE}}^{(k)}(\hat{\thvec}^{(k)},\thvecvar^{(k)})  \succeq {\bf{MCRB}}^{(k)}(\thvecvar^{(k)}), 
			\ee
		where  the  $k$th-PS-MCRB is given by
			\be \label{MCRB_k} \begin{aligned}[b]
			{\bf{MCRB}}^{(k)}&(\thvecvar^{(k)}) \\
    \triangleq& (\Amat^{(k)}(\thvecvar^{(k)}))^{-1}{\Bmat}^{(k)}(\thvecvar^{(k)})(\Amat^{(k)}(\thvecvar^{(k)}))^{-1},
			\end{aligned}
			\ee
			  the pseudo-true parameter vector $\thvecvar^{(k)}$  and the  post-model-selection Hessian form information matrix, $\Amat^{(k)}(\thvec^{(k)})$, are defined in 
\eqref{k_pseudo_true_parameter_kld} and \eqref{Amat}, respectively, and the outer-product form of the $k$th post-model-selection  information matrix is
			\be \label{Bmat}
			\Bmat^{(k)}(\thvec^{(k)})\triangleq\mathrm{E}_p\left[ \grad{\thvec^{(k)}}\log f(\xvec;\thvec) \grad{\thvec^{(k)}}^T\log f(\xvec;\thvec)  | \Psi=k \right].
			\ee
		\end{Theorem}
	  \begin{IEEEproof}
	  	The proof of \cref{PS_MCRB_k}
    can be obtained along the lines of 
     the proof of the MCRB for the conventional case from
	  	\cite[Th. 4.1]{vuong1986cramer}, by replacing the true pdf (denoted in \cite{vuong1986cramer} as $g(\cdot,\cdot)$) with the true conditional pdf from \eqref {p_given_psi}, $p(\xvec|\Psi=k;\bvarphi)$, and replacing the estimator (denoted in \cite{vuong1986cramer} as $T(\cdot)$) with an estimator of the $k$th parameter vector, $\hat{\thvec}^{(k)}$. As a result, all the expectations involved in the terms ${\bf{MSE}}^{(k)}(\hat{\thvec}^{(k)},\thvecvar^{(k)})$, $\Amat^{(k)}(\thvecvar^{(k)})$, and ${\Bmat}^{(k)}(\thvecvar^{(k)})$ are conditional expectations that are computed using integration w.r.t. 
        $p(\xvec|\Psi=k;\bvarphi)$. 
	  \end{IEEEproof}

   The bound in Theorem \ref{PS_MCRB_k} provides a marginal lower bound on each $k$th-MSE from \eqref{MSE_k}. 	In general, this  $k$th-MSE matrix may have different dimensions for every $k$, $|\Omega_k|\times |\Omega_k|$, where  $|\Omega_k|$ is the dimensions of the $k$th parameter vector,  $\thvec^{(k)}$. 
    Thus, in order to derive a total bound,  a general estimation task across all models should be considered.

 We first remind that $\bvarphi \in \Omega_{\varphi}$ is the general parameter vector of interest, as defined in \cref{PMS_est}.
   We assume that
     for any candidate model, there is a deterministic, continuously differentiable 
	  	mapping from the $k$th model parameter space to $\Omega_{\varphi}$, represented by $\bvarphi_k: \Omega_{k}\rightarrow \Omega_{\varphi} $, such that 
  $\bvarphi_k(\thvec^{(k)})=  \bvarphi$.
	  	Therefore, any practical post-model-selection estimator of $\bvarphi$, $\hat{\bvarphi}: \Omega_{\xvec}\rightarrow \Omega_{\varphi}$, can be written in the following form:
	  	\be \label{generic_post_selection_estimator}
	  	\hat{\bvarphi}=\sum\nolimits_{k=1}^{K}\bvarphi_k(\hat{\thvec}^{(k)})\ind{\xvec\in\Am{k}}.
	  	\ee
      The MSE of an  estimator   $\hat{\bvarphi}$ from  \eqref{generic_post_selection_estimator} is  defined as
	 	\be \label{MSE}
	 {	\bf{MSE}}(\hat{\bvarphi},\bvarphi)\triangleq \mathrm{E}_p\left[  (\hat{\bvarphi}-\bvarphi)(\hat{\bvarphi}-\bvarphi)^T  \right].
	 	\ee

   In order to incorporate the model selection stage and analyze the estimators under each selected model separately, we  decompose the MSE from \eqref{MSE} by  substituting a general 
   estimator $\hat{\bvarphi}$ 
   from \eqref{generic_post_selection_estimator} 
   and using the law of total expectation:
	 	\be \label{MSE_phi}\begin{aligned}[b]
	 			{\bf{MSE}}(&\hat{\bvarphi},\bvarphi)=
	 			\\ \sum_{k=1}^K p_k&\mathrm{E}_p\left[  (\bvarphi_k(\hat{\thvec}^{(k)})-\bvarphi)(\bvarphi_k(\hat{\thvec}^{(k)})-\bvarphi)^T |\Psi=k \right],
	 	\end{aligned}
	 		 	\ee
      where $p_k$ is defined in \eqref{p_k}, and is a function of $\bvarphi$.
	The conditional expectations in \eqref{MSE_phi} are calculated w.r.t. the conditional pdfs in \eqref{p_given_psi}.
   The following Theorem uses \cref{PS_MCRB_k} to obtain the PS-MCRB on the total MSE from \eqref{MSE}.
		\begin{Theorem}(PS-MCRB) \label{PS_MCRB}
                Let us assume 
                a post-model-selection model with the assumed pdf, $f(\xvec;\thvec)$, and a selection rule $\Psi$ that satisfy \cref{cond1,cond2,cond3,cond4}, $\forall k \in \{1,\ldots,K\}$. The MSE of any post-model-selection estimator $\hat{\bvarphi}$ in \eqref{generic_post_selection_estimator}, which is consisted by PSMS-unbiased for all $k \in \{1,\ldots,K\}$, satisfies
                \be \label{PS_MCRB_q}\begin{aligned}[b]
				{\bf{MSE}}(&\hat{\bvarphi},\bvarphi)
					\succeq \sum_{k=1}^Kp_k\left(\dot{\bvarphi_k}^T(\thvecvar^{(k)}){\bf MCRB}^{(k)}\dot{\bvarphi_k}(\thvecvar^{(k)}) \right. \\
				& \left.+(\bvarphi_k(\thvecvar^{(k)})-\bvarphi)(\bvarphi_k(\thvecvar^{(k)})-\bvarphi)^T\right),
			\end{aligned}
			\ee
                where
			\be
				\dot{\bvarphi_k}(\thvecvar^{(k)}) \triangleq \left.\grad{\thvec^{(k)}} \bvarphi_k^T({\thvec}^{(k)}) \right|_{\thvec^{(k)}=\thvecvar^{(k)}}
			\ee
			is the Jacobian matrix of the $k$th mapping $\bvarphi_k(\cdot)$ evaluated at the relevant pseudo-true parameter vector, $\thvecvar^{(k)}$.
        \end{Theorem}

        \begin{IEEEproof}
                Since we assumed that for any $k$, $\bvarphi_k(\cdot)$, is a continuously differentiable mapping, similar to the conventional CRB on a 
 functional
transformation of the unknown parameter vector 
                (see e.g. \cite{vuong1986cramer}, \cite{kay1993fundamentals}), the marginal bound in 
                \eqref{bound1} can be generalized to the estimation of 
			\be \label{MCRB_k_wrt_phi}\begin{aligned}[b]
				\mathrm{E}_p&\left[  (\bvarphi_k(\hat{\thvec}^{(k)})-\bvarphi)(\bvarphi_k(\hat{\thvec}^{(k)})-\bvarphi)^T |\Psi=k \right]
				\\	\succeq& \dot{\bvarphi_k}^T(\thvecvar^{(k)}){\bf MCRB}^{(k)}\dot{\bvarphi_k}(\thvecvar^{(k)})\\
				&+(\bvarphi_k(\thvecvar^{(k)})-\bvarphi)(\bvarphi_k(\thvecvar^{(k)})-\bvarphi)^T,
			\end{aligned}
			\ee
			where the last term in \eqref{MCRB_k_wrt_phi} is
   since the transformation of the pseudo-true parameter vector, $\bvarphi_k(\thvecvar^{(k)})$, is not necessarily equal to $\bvarphi$.
			By plugging the marginal bounds from \cref{MCRB_k_wrt_phi} for  $k=1,\ldots,K$ in \cref{MSE_phi}, and since the probabilities of selection, $\{p_k\}_{k=1}^K$ are non-negative,
   we obtain 
   \eqref{PS_MCRB_q}.
		 \end{IEEEproof}

     In \cref{PS_MCRB_k,PS_MCRB}, we derived the $k$th-PS-MCRB and PS-MCRB for a general assumed pdf, $f(\xvec;\thvec)$. In the following, we  implement the associated post-model-selection information matrices  $\Amat^{(k)}(\cdot)$ and
			$\Bmat^{(k)}(\cdot)$ from \eqref{Amat} and 
			\eqref{Bmat}  for the three interpretation from  \cref{relation_misspecified}. 
 The different post-model-selection information matrices lead to different MCRBs. 

				\subsubsection{Naive Interpretation}
            Although $f_I(\xvec;\thvec)$ is not a valid pdf (see \cref{Naive_Interpretation}), since 
             from \eqref{f_I} $f_I(\xvec;\thvec)=f_k(\xvec;\thvec^{(k)})$, $\forall\xvec \in \Am{k}$, we  require \cref{cond1,cond2,cond3,cond4} to be satisfied for $f_k(\xvec;\thvec^{(k)})$, and  then \cref{PS_MCRB} is applied for $f_I(\xvec;\thvec)=f_k(\xvec;\thvec^{(k)})$ under the $k$th selection.
             In addition, 
				\be \label{grad_log_f_I}
					\grad{\thvec^{(k)}} \log f_I(\xvec;\thvec)=\grad{\thvec^{(k)}} \log f_k(\xvec;\thvec^{(k)}), ~~\forall \xvec\in\Am{k},
				\ee
    $ k\in \{1,\ldots, K\}$. 
				By substituting \eqref{grad_log_f_I} in \cref{Amat} and  \cref{Bmat}
    we obtain that the $k$th post-model-selection
    Hessian form and outer-product form information matrices, in this case, are
				\be \label{Ak_I}
					\Amat^{(k)}_I(\thvec^{(k)}) = \mathrm{E}_p\left[ \grad{\thvec^{(k)}}^2 \log f_k(\xvec;\thvec^{(k)}) | \Psi=k  \right]
				\ee
				and
	\be\label{Bk_I} \begin{aligned}[b]
	&\Bmat^{(k)}_I(\thvec^{(k)})=
	\\&\mathrm{E}_p \left[ \grad{\thvec^{(k)}}\log f_k(\xvec;\thvec^{(k)}) \grad{\thvec^{(k)}}^T\log f_k(\xvec;\thvec^{(k)})  | \Psi=k \right].
		\end{aligned} 
\ee
				
				\subsubsection{Normalized Interpretation}
					To obtain the $k$th post-model-selection information matrices under the normalized interpretation, we use the derivative of $\log f_{II}(\xvec;\thvec)$ from \eqref{f_II}:
				\begin{eqnarray} \label{grad_log_f_II} 
    	\grad{\thvec^{(k)}} \log f_{II}(\xvec;\thvec)\hspace{4.75cm}\nonumber\\=\grad{\thvec^{(k)}} \log f_k(\xvec;\thvec^{(k)})-\grad{\thvec^{(k)}} \log \alpha(\thvec), ~~\forall \xvec\in\Am{k},
				\end{eqnarray}
    $ k=1,\ldots, K$. 
				By substituting \eqref{grad_log_f_II} in \cref{Amat} we obtain that
				\be \label{Ak_II}\begin{aligned}[b]
					\Amat^{(k)}_{II}(\thvec) = 
     \Amat^{(k)}_I(\thvec^{(k)})
					-\grad{\thvec^{(k)}} ^2 \log \alpha(\thvec),
				\end{aligned}
				\ee
     where $ \Amat^{(k)}_I(\cdot)$ is defined in \eqref{Ak_I}. 
				Similarly, by substituting \eqref{grad_log_f_II} in \cref{Bmat} 
     and using the fact that $\grad{\thvec^{(k)}} \log \alpha(\thvec)$ is deterministic, 
    we obtain that
				\be \label{Bmat_IIa}\begin{aligned}[b]
					\Bmat&^{(k)}_{II}(\thvec^{(k)})\hspace{-0.1cm}=\Bmat^{(k)}_I(\thvec^{(k)})\\
					&-\mathrm{E}_p\hspace{-0.1cm}\left[ \grad{\thvec^{(k)}}\log f_k(\xvec;\thvec)  | \Psi=k \right]\grad{\thvec^{(k)}}^T \log \alpha(\thvec)\\
					&-\grad{\thvec^{(k)}} \log \alpha(\thvec)\mathrm{E}_p\hspace{-0.1cm}\left[ \grad{\thvec^{(k)}}^T\log f_k(\xvec;\thvec)  | \Psi=k \right]
					\\&+\grad{\thvec^{(k)}}\log \alpha(\thvec) \grad{\thvec^{(k)}}^T \log \alpha(\thvec),
				\end{aligned}
				\ee
        where $\Bmat^{(k)}_I(\cdot)$ is defined in \eqref{Bk_I}.
			Since the pseudo-true parameter vector under this interpretation, $\thvecvar_{II}$, 
  maximizes the r.h.s. of \eqref{k_pt_II}, then under 
    regularity condition \ref{cond2} for this case (i.e. twice differentiability of $f_{II}(\xvec;\thvec)$), 
   $\thvecvar^{(k)}_{II}$ is a stationary point that satisfy 
    	$\grad{\thvec^{(k)}} \log f_{II}(\xvec;\thvec)=\zerovec$, which by \eqref{grad_log_f_II} implies that
			\be
   \label{pseudo_II}
				\grad{\thvec^{(k)}} \log \alpha(\thvecvar_{II})=\left.\mathrm{E}_p\hspace{-0.1cm}\left[ \grad{\thvec^{(k)}}\log f_k(\xvec;\thvec)  | \Psi=k \right]\right|_{\thvec^{(k)}=\thvecvar^{(k)}_{II}},
			\ee
			$\forall k \in \{1,\ldots,K\}$. By substituting \eqref{pseudo_II} in \eqref{Bmat_IIa} we obtain that 
				\be \label{Bk_II}\begin{aligned}[b]
				\Bmat^{(k)}_{II}(\thvecvar^{(k)}_{II})\hspace{-0.1cm}=\Bmat^{(k)}_I(\thvecvar^{(k)}_{II})
				-\grad{\thvec^{(k)}}\log \alpha(\thvecvar_{II}) \grad{\thvec^{(k)}}^T \log \alpha(\thvecvar_{II}).
			\end{aligned}
			\ee

   It can be seen that $\Amat^{(k)}_{II}(\cdot)$ and $\Bmat^{(k)}_{II}(\cdot)$ from \eqref{Ak_II} and \eqref{Bk_II}, respectively, are both composed of the sum of the $k$th post-model-selection matrices 
   of the naive interpretation, $\Amat^{(k)}_{I}(\cdot)$ and $\Bmat^{(k)}_{I}(\cdot)$, and a second term  which stems from the factor $\alpha(\thvec)$ from \eqref{alpha}  and is determined by the selection approach.
  However, the PS-MCRB under this normalized interpretation is based on evaluating  $\Amat^{(k)}_{II}(\cdot)$ and $\Bmat^{(k)}_{II}(\cdot)$ at the pseudo-true parameter vector under this interpretation, $\thvecvar_{II}$, and not at $\thvecvar_{I}$.
				\subsubsection{Selective Inference Interpretation}    \label{bound_III_subsec}
					To obtain the $k$th post-model-selection information matrices under the selective inference interpretation,  we use the derivative of $\log f_{III}(\xvec;\thvec)$ from \cref{f_III_2} (using the fact that $c_k$ is not a function of $\xvec$):
				\be \label{grad_log_f_III} \begin{aligned}[b]
						\grad{\thvec^{(k)}} \log &f_{III}(\xvec;\thvec)=
						\\ &\grad{\thvec^{(k)}} \log f_k(\xvec;\thvec^{(k)})-\grad{\thvec^{(k)}} \log \pi_k(\thvec^{(k)}_{III}),
				\end{aligned}
				\ee
    for any $\xvec\in\Am{k}$.
				By substituting \eqref{grad_log_f_III} in \cref{Amat} we obtain that
				\be \label{Ak_III}\begin{aligned}[b]
					\Amat^{(k)}_{III}(\thvecvar^{(k)}_{III}) =  \Amat^{(k)}_I(\thvecvar^{(k)}_{III})
				-\grad{\thvec^{(k)}} ^2 \log \pi_k(\thvecvar^{(k)}_{III}),
				\end{aligned}
				\ee
    where $ \Amat^{(k)}_I(\cdot)$ is defined in \eqref{Ak_I}.
				Similarly, by substituting the gradient from \eqref{grad_log_f_III} in \eqref{Bmat} and using the fact that $\grad{\thvec^{(k)}} \log \pi_k(\thvecvar^{(k)}_{III})$ is deterministic,  we obtain that
				\be \label{Bmat_IIIa}\begin{aligned}[b]
					\Bmat&^{(k)}_{III}(\thvec^{(k)})
					=\Bmat^{(k)}_I(\thvec^{(k)})
					\\&-\mathrm{E}_p\left[ \grad{\thvec^{(k)}}\log f_k(\xvec;\thvec^{(k)})  | \Psi=k \right]\grad{\thvec^{(k)}}^T \log \pi_k(\thvec^{(k)})\\
					&-\grad{\thvec^{(k)}} \log \pi_k(\thvec^{(k)})\mathrm{E}_p\hspace{-0.1cm}\left[ \grad{\thvec^{(k)}}^T\log f_k(\xvec;\thvec)  | \Psi=k \right]
					\\&+\grad{\thvec^{(k)}}\log \pi_k(\thvec^{(k)}) \grad{\thvec^{(k)}}^T \log \pi_k(\thvec^{(k)}),
				\end{aligned}
				\ee
    where $	\Bmat^{(k)}_I(\cdot)$ is defined in \eqref{Bk_I}.
				Since the pseudo-true parameter vector,  $\thvecvar^{(k)}_{III}$, 
    maximizes the r.h.s. of \eqref{k_pt_III}, then under 
    regularity condition \ref{cond2} (i.e. twice differentiability of $f_{III}(\xvec;\thvec)$), 
   $\thvecvar^{(k)}_{III}$ is a stationary point that 
  satisfy 
    	$\grad{\thvec^{(k)}} \log f_{III}(\xvec;\thvec)=\zerovec$, which by \eqref{grad_log_f_III} implies that
				\be \label{pseudo_III}
				\grad{\thvec^{(k)}} \log \pi_k(\thvecvar^{(k)}_{III})=
				\left.\mathrm{E}_p\left[ \grad{\thvec^{(k)}}\log f_k(\xvec;\thvec)  | \Psi=k \right]\right|_{\thvec^{(k)}=\thvecvar^{(k)}_{III}},
				\ee
    	$\forall k \in \{1,\ldots,K\}$.
				By substituting \eqref{pseudo_III} in \eqref{Bmat_IIIa} we obtain that 
					\be \label{Bk_III}\begin{aligned}[b]
					\Bmat^{(k)}_{III}(\thvecvar^{(k)}_{III})&=\Bmat^{(k)}_I(\thvecvar^{(k)}_{III})\\
				&-\grad{\thvec^{(k)}}\log \pi_k(\thvecvar^{(k)}_{III}) \grad{\thvec^{(k)}}^T \log \pi_k(\thvecvar^{(k)}_{III}).
				\end{aligned}
				\ee
    
  It can be seen that  $\Amat^{(k)}_{III}
  (\cdot)$ and $\Bmat^{(k)}_{III} (\cdot)$ from \eqref{Ak_III} and  \eqref{Bk_III}, respectively, 
are composed of the sum of the $k$th post-model-selection information matrices of the naive interpretation, $\Amat^{(k)}_{I}(\cdot)$ and $\Bmat^{(k)}_{I}(\cdot)$ from   \eqref{Ak_I} and \eqref{Bk_I}, and a second term which is determined by the model selection, based on derivatives of  $\pi_k(\cdot)$.  All these terms are evaluated at the associated pseudo-true parameter vector, $\thvecvar^{(k)}_{III}$.
	\begin{remark} \label{single_candidate_MCRB}
	For the special case of a single candidate model, i.e. if $K=1$, the post-model-selection pseudo true parameter vector from \cref{k_pseudo_true} coincides with the conventional pseudo true parameter vector from \cref{Pseudo_true}. Moreover, the PS-MCRB coincides with the conventional MCRB 
 described in \cref{conventional_MCRB}.
	\end{remark}	
		\begin{remark}
    The $k$th-MSE in \eqref{MSE_k__} is defined w.r.t. the $k$th pseudo-true parameter vector.
    Since each interpretation has different pseudo-true parameter vectors, the $k$th-MSE is defined differently for each interpretation. 
    Thus, the $k$th-PS-MCRB under each of the interpretations is a lower bound on a different risk. 
     Moreover, according to  \cref{PSMS_unbiasedness_def}, each interpretation induces a different PSMS-unbiasedness condition. Hence,
     each interpretation results in a bound for a different class of estimators. 
     \end{remark}

 \begin{remark}
 Under mild regularity conditions \cite{kay1993fundamentals} and if  $\hat{\bvarphi}$ is a mean-unbiased estimator of $\bvarphi$, its MSE is bounded by the following oracle CRB:
	 	\be
   \label{oracle_crb1}
   {\bf{MSE}}(\hat{\bvarphi},\bvarphi)\succeq 
\mathrm{E}_p[\grad{\bvarphismall}\log p(\xvec;\bvarphi)\grad{\bvarphismall}^T\log p(\xvec;\bvarphi)].
	 	\ee
The oracle CRB in \eqref{oracle_crb1}  is	commonly used in post-model-selection estimation analysis \cite{ye2003cramer,meir2021cramer,ben2010cramer}. 	 
However, it is not a valid bound since it relies on knowledge of the true model, disregards the selection stage, and does not account for model misspecification, which impacts the estimation performance.
Thus, it is only used here as a theoretical benchmark.
 \end{remark}

	 	\section{Example: Estimation After Detection}\label{simulations}
	 	
	 	Let $\xvec \in \mathbb{R}^N$ be an observation vector such that
	 	\be
   \label{model_example}
	 	\xvec =\bvarphi +\wvec,
	 	\ee
	 	where $\bvarphi \in \mathbb{R}^N$ is an unknown deterministic parameter vector to be estimated, and $\wvec$ is white Gaussian noise with zero mean and a known covariance matrix, $\sigma^2\Imat$.
	 	We consider two hypotheses regarding $\bvarphi$:
	 	\be \label{hypotheses}
	 	\begin{cases}
	 		{\cal H}_1: \bvarphi=\Hmat\thvec^{(1)}\\
	 		{\cal H}_2: \bvarphi = \thvec^{(2)},
	 	\end{cases}
	 	\ee
	 	i.e. under ${\cal H}_1$, the unknown parameter vector $\bvarphi \in \mathbb{R}^{N}$ belongs to the column space of $\Hmat \in \mathbb{R}^{N\times M}$, which is a known rank  $M$ matrix, where $M<N$. 
   Under ${\cal H}_2$, $\bvarphi \in \mathbb{R}^{N}$ does not have a specific structure.
	 	Thus, the unknown parameter vector has a different dimension under each hypothesis (aka model): $\thvec^{(1)} \in \mathbb{R}^M$ and $\thvec^{(2)} \in \mathbb{R}^N$.
   For example, in communication systems, ${\cal H}_1$ can describe a scenario where the signal is received from a known channel, $\Hmat$, while hypothesis ${\cal H}_2$ describes the model of 
   signal received from an unknown channel. 
   In the considered setting,  the candidate pdfs are both Gaussian with means $\Hmat\thvec^{(1)}$ and $\thvec^{(2)}$, respectively.
	 		 	
	 	Based on the observation vector $\xvec$, one of the hypotheses/models is selected. We consider here the following generalized likelihood ratio test  (GLRT) as the selection rule \cite{kay1998fundamentals
   }:
   \be
    \label{Psi_GLRT2}
        \Psi=\begin{cases}
	 		1, &\frac{1}{\sigma^2}\xvec^T\Pmat_\Hmat^\bot\xvec \leq \gamma\\
	 		2, &{\text{otherwise}},
	 	\end{cases}
   \ee
   	 	where
            $ 	\Pmat_\Hmat^\bot \triangleq \Imat - \Hmat(\Hmat^T\Hmat)^{-1}\Hmat^T$. 
   
 In this case, 
 any practical post-model-selection estimator of  the parameter vector of interest  $\bvarphi$ has the following form
	 	\be \begin{aligned}[b]
   \hat{\bvarphi} = &
    \hat{\bvarphi}^{(1)}\ind{\xvec\in\Am{1}} +
      \hat{\bvarphi}^{(2)}\ind{\xvec\in\Am{2}}\\
    =&
    \Hmat\hat{\thvec}^{(1)}\ind{\xvec\in\Am{1}} + \hat{\thvec}^{(2)}\ind{\xvec\in\Am{2}},	    
	 	\end{aligned}
	 	\ee
   where $\hat{\thvec}^{(k)}$, $k=1,2$ are the estimators of the unknown parameter vector under hypothesis $k=1,2$, and the events $\xvec\in \Am{1}$ and $ \xvec\in\Am{2}$ are determined according to \eqref{Psi_GLRT2}. 
	 	
	 Under the considered settings, $\frac{1}{\sigma^2}\xvec^T\Pmat_\Hmat^\bot\xvec 
  $
  has a $\chi^2$ distribution with $
	 r= \text{Rank}(\Pmat_\Hmat^\bot)=N-M
	 $ degrees of freedom, and non-centrality parameter \cite[Ch. 2.3]{kay1998fundamentals}
	 \be \label{lambda}
	 \lambda \triangleq  \frac{\bvarphi^T\Pmat_\Hmat^\bot\bvarphi}{\sigma^2}.
	 \ee
	 Thus, the true probabilities of selection from \eqref{p_k} in this case are
	 \be \label{pk_chi2}
	 	p_k=\begin{cases}
	 		 F_r( \gamma; \lambda), &k=1 \\
	 		1- F_r( \gamma; \lambda), &k=2,
	 	\end{cases}
	 \ee
	 where $F_r(\cdot;\lambda)$ is the $\chi^2$ cdf with $r$ degrees of freedom and non-centrality parameter $\lambda$.
Since
$\Pmat_\Hmat^\bot\Hmat=\zerovec$, under ${\cal H}_1$ the non-centrality in \cref{lambda} vanishes and the test has a central $\chi^2$ distribution $\forall \thvec^{(1)}  \in \mathbb{R}^M$. Under ${\cal H}_2$ the non-centrality is given by
 	\be \label{lambda_2}
	 \lambda^{(2)}=\frac{(\thvec^{(2)})^T\Pmat_\Hmat^\bot\thvec^{(2)}}{\sigma^2}.
	 \ee 
Therefore, the assumed probabilities of selection from \eqref{pi_k} are given by
	 \be \label{pi_k_chi2}
	 \pi_k(\thvec^{(k)})=\begin{cases}
	 	F_r( \gamma; 0), &k=1 \\
	 	1- F_r( \gamma; \lambda^{(2)}), &k=2.
	 \end{cases}
	 \ee



	 	In the following, we present the estimators and bounds under the three interpretations for this scenario. Detailed derivations appear in the supplemental materials in \cite{harel2023_misspecified_SupplementaryMaterials}.
	 	\subsection{Estimators} \label{supp_estimators}
	 	In this subsection, we present the estimators from \cref{estimators}, according to each interpretation, for this example.
	 	\subsubsection{Naive Interpretation} \label{supp_estimators_I}
	 	Since $\{f_k(\xvec;\thvec^{(k)}\}_{k=1}^2$, are Gaussian pdfs with the means from 
  \eqref{model_example}, \eqref{hypotheses}, the MSL estimator from \eqref{MSL_2} is given by \cite[\cref{supp_MSL_estimator}]{harel2023_misspecified_SupplementaryMaterials}
	 	\be 
   \label{msl_ex1}
	 		\msl{\thvec}^{(k)}
	 		=\begin{cases}
	 			(\Hmat^T\Hmat)^{-1}\Hmat^T\xvec, &k=1 \\
	 			\xvec, &k=2.
	 		\end{cases}
	 	\ee
	 	\subsubsection{Normalized Interpretation} \label{supp_estimators_II}
	Since $\pi_1$ is not a function of $\thvec^{(1)}$, $\alpha(\thvec)$ is not a function of $\thvec^{(1)}$.
	Thus, the maximization in \eqref{MSNL_est1} for $k=1$ w.r.t. $\thvec^{(1)}$ is equivalent to the maximization in \eqref{MSL_2}. Therefore, if $\xvec \in \Am{1}$, the MSNL estimator is 
	 	\be 
   \label{msnl_ex1}
	 	\msnl{\thvec}^{(1)}= (\Hmat^T\Hmat)^{-1}\Hmat^T\xvec.
	 	\ee
	 	If  $\xvec \in \Am{2}$, the MSNL estimator is obtained by the maximization in \eqref{MSNL_est1} w.r.t. $\thvec^{(2)}$, which is obtained by setting the gradient of the r.h.s. of \eqref{MSNL_est1}  to zero.
	 	In \cite[\cref{supp_MSNL_estimator}]{harel2023_misspecified_SupplementaryMaterials}, we show that this results in the following score equation:
	 	\be \label{MSNL_score} 
	\xvec-\thvec^{(2)}- \frac{F_r(\gamma; \lambda^{(2)} )-F_{r+2}(\gamma; \lambda^{(2)}
		) }{F_r( \gamma; 0)+1-F_r(\gamma;  \lambda^{(2)})}\Pmat_\Hmat^\bot\thvec^{(2)}=\zerovec.
	\ee
	 Then, we set the MSNL estimator, $\msnl{\thvec}^{(2)}$, to be the solution of \eqref{MSNL_score}, which can be found numerically. 
	 	\subsubsection{Selective Inference Interpretation} \label{supp_estimators_III}
	 	The PSML from \cref{PSML} is given by the maximization in \cref{PSML_est_k_log}. 
   Since 
   $\pi_1(\thvec^{(1)})$ is 
   not a function of $\thvec^{(1)}$, the maximization in \eqref{PSML_est_k_log} w.r.t. $\thvec^{(1)}$ is equivalent to the maximization in \eqref{MSL_2}, as before. Thus, 
  if ${\cal H}_1$ is selected, the PSML estimator of $\thvec^{(1)}$ is 
	 	\be \label{PSML_est0} 
	 	\psml{\thvec}^{(1)}=(\Hmat^T\Hmat)^{-1}\Hmat^T\xvec,
	 	\ee
   which coincides with the MSL and MSNL estimators of $\thvec^{(1)}$ from \eqref{msl_ex1} and \eqref{msnl_ex1}, respectively. If ${\cal H}_2$ is selected, the PSML estimator of $\thvec^{(2)}$ is obtained by the maximization of \eqref{PSML_est_k_log} w.r.t. $\thvec^{(2)}$. This maximization is obtained by setting the gradient of the r.h.s. of \eqref{PSML_est_k_log} (with $
	 \pi_k(\thvec^{(k)})$ from \eqref{pi_k_chi2} and the Gaussian pdf $f_2(\xvec;\thvec^{(2)})$) to zero, which results in 
  \cite[\cref{supp_PSML_estimator}]{harel2023_misspecified_SupplementaryMaterials}
	 \be \label{PSML_score} 
	\xvec-\thvec^{(2)}
	- \frac{(F_r(\gamma; \lambda^{(2)} )-F_{r+2}(\gamma; \lambda^{(2)} )) }{1-F_r(\gamma; \lambda^{(2)})}\Pmat_\Hmat^\bot\thvec^{(2)}
	=\zerovec.
	\ee
 Then, we set the PSML estimator, $\psml{\thvec}^{(2)}$, to be the solution of \eqref{PSML_score}, which can be found by  numerical method. 
	 \vspace{-0.25cm}
	 	 \subsection{Pseudo-true Parameter Vectors} \label{example_PT}
	 	 In this subsection, we present the pseudo-true parameter vectors from \cref{pseudo_true_section}, according to each interpretation.
        The full derivation appears in \cite[\cref{supp_PT}]{harel2023_misspecified_SupplementaryMaterials}.
	 	 \subsubsection{Naive Interpretation} \label{supp_PT_I}
	 	 By substituting the considered setting in \cref{k_pt_I}, we show in \cite[\cref{supp_PT_naive}]{harel2023_misspecified_SupplementaryMaterials} that the pseudo-true parameter vector according to the naive interpretation:
	 	 \be \label{vecvar_I}
	 	 {\thvecvar}^{(k)}_{I}
	 	 =\begin{cases}
	 	 	(\Hmat^T\Hmat)^{-1}\Hmat^T \muvec^{(1)}, &k=1 \\
	 	 	\muvec^{(2)}, &k=2,
	 	 \end{cases}
	 	 \ee
	 	 where the conditional expectation of $\xvec$ given $\Psi=k$ is
	 	\be
   \label{mu_k}
	 	 \muvec^{(k)}\triangleq \mathrm{E}_p \left[ \xvec|\Psi=k \right],~~~k \in{1,2},
            \ee
	 	which has a   closed-form expression  (see in \cite[\cref{cond_expectations}]{harel2023_misspecified_SupplementaryMaterials}).
	 	 \subsubsection{Normalized Interpretation} \label{supp_PT_II}
	 	 Since in this case $\pi_1(\thvec^{(1)})$ is not a function of $\thvec^{(1)}$, the gradient of \cref{k_pt_II_final} w.r.t. $\thvec^{(1)}$ is as in the naive interpretation. Therefore, the pseudo-true parameter vector according to the normalized interpretation under ${\cal H}_1$ is 
	 	 \be \label{vecvar_1_II}
	 	 {\thvecvar}^{(1)}_{II}={\thvecvar}^{(1)}_{I}=(\Hmat^T\Hmat)^{-1}\Hmat^T \muvec^{(1)}.
	 	 \ee
	 	 The pseudo-true parameter vector under ${\cal H}_2$ is obtained by
	 	  setting the gradient of \eqref{k_pt_II_final} w.r.t. $\thvec^{(2)}$ to zero, which results in
	 	  \be \label{vecvar_2_II}
	 	  \muvec^{(2)}-\thvec^{(2)}- \frac{(F_r(\gamma; \lambda^{(2)} )-F_{r+2}(\gamma; \lambda^{(2)} )) }{\alpha(\thvec)}\Pmat_\Hmat^\bot\thvec^{(2)}=\zerovec,
	 	  \ee
	  which can be solved numerically.
	 	 \subsubsection{Selective Inference Interpretation} \label{supp_PT_III}
	 	 By substituting the considered setting in \eqref{k_pt_III},  we have
	 	\be \label{vecvar_1_III}
	 	{\thvecvar}^{(1)}_{III}={\thvecvar}^{(1)}_{I}=(\Hmat^T\Hmat)^{-1}\Hmat^T \muvec^{(1)}.
	 	\ee
	By setting the gradient of \eqref{k_pt_III} w.r.t. $\thvec^{(2)}$ to zero, we obtain that the pseudo-true parameter vector under ${\cal H}_2$ ($\Psi=2$) is \cite{harel2023_misspecified_SupplementaryMaterials}:
  
  \be
  \label{vecvar_III}
 {\thvecvar}^{(2)}_{III}=\bvarphi.
 \ee
	 	 \subsection{PS-MCRBs}
	 	  In this subsection, we present the  PS-MCRBs from \cref{bounds}, according to each interpretation.
   The Hessian matrix of the log-likelihood under each of the hypotheses is given by
	 	 \be \label{example1_J_A}
	 	 \grad{\thvec^{(k)}}^2 \log f_k(\xvec;\thvec^{(k)})=\begin{cases}
	 	 	-\frac{1}{\sigma^2}\Hmat^T\Hmat, &k=1 \\
	 	 	-\frac{1}{\sigma^2}\Imat, &k=2, 
	 	 \end{cases}
	 	 \ee
    which is a deterministic matrix independent of $\xvec$. Thus, 
    by substituting \eqref{example1_J_A} in \cref{Ak_I}, \cref{Ak_II}, and \cref{Ak_III}, we obtain the  post-model-selection Hessian form information matrices:
		\be \label{example_Ak_I}
		\Amat_{I}^{(k)}(\thvecvar^{(k)}_{I})=\begin{cases}
			-\frac{1}{\sigma^2}\Hmat^T\Hmat, &k=1 \\
			-\frac{1}{\sigma^2}\Imat, &k=2,
		\end{cases}
		\ee
		\be \label{example_Ak_II}
		\Amat_{II}^{(k)}(\thvecvar^{(k)}_{II})=\begin{cases}
			-\frac{1}{\sigma^2}\Hmat^T\Hmat, &k=1 \\
			-\frac{1}{\sigma^2}\Imat-\grad{\thvec^{(2)}} ^2 \log \alpha(\thvecvar_{II}), &k=2,
		\end{cases}
		\ee
  and
		\be \label{example_Ak_III}
		\Amat_{III}^{(k)}(\thvecvar^{(k)}_{III})=\begin{cases}
			-\frac{1}{\sigma^2}\Hmat^T\Hmat, &k=1 \\
			-\frac{1}{\sigma^2}\Imat-\grad{\thvec^{(2)}} ^2 \log \pi_2(\thvecvar^{(2)}_{III}), &k=2,
		\end{cases}
		\ee
  respectively.
			Closed-form  expressions of $\grad{\thvec^{(2)}} ^2 \log\pi_2(\thvec)$  and $\grad{\thvec^{(2)}} ^2 \log \alpha(\thvec)$ appear in \cite[\cref{supp_PSMCRB_II,supp_PSMCRB_III}]{harel2023_misspecified_SupplementaryMaterials}.

	 	 In addition, in \cite[\cref{supp_PS_MCRB}]{harel2023_misspecified_SupplementaryMaterials} it is shown that 
    by substituting the considered settings and the associated pseudo-true parameter for each interpretation from Subsection \ref{supp_PT}  in \cref{Bk_I,Bk_II,Bk_III},
 the  outer-product form of the $k$th post-model-selection  information matrices are identical:
\be \label{outer_FIM_example}
	\Bmat_{i}^{(k)}(\thvecvar^{(k)}_i)=\begin{cases}
				\frac{1}{\sigma^4}\Hmat^T\bsigma_{\xvec}^{(1)}\Hmat, &k=1 \\
				\frac{1}{\sigma^4}\bsigma_{\xvec}^{(2)}, &k=2,
			\end{cases}
\ee
where $i= I,II,III$ and
\be \label{cov_k}
\bsigma_{\xvec}^{(k)} \triangleq \mathrm{E}_p [ (\xvec- \muvec^{(k)})(\xvec- \muvec^{(k)})^T|\Psi=k ],~ k \in{1,2}.
\ee
A closed-form expression of $\bsigma_{\xvec}^{(k)}$ is given in \cite[\cref{cond_expectations}]{harel2023_misspecified_SupplementaryMaterials}.
\vspace{-0.5cm}
	\subsection{Simulation Results}
\begin{figure}[htb] 
	 			\centering\vspace{-0.25cm}
	 			\includegraphics[width=6.75cm]{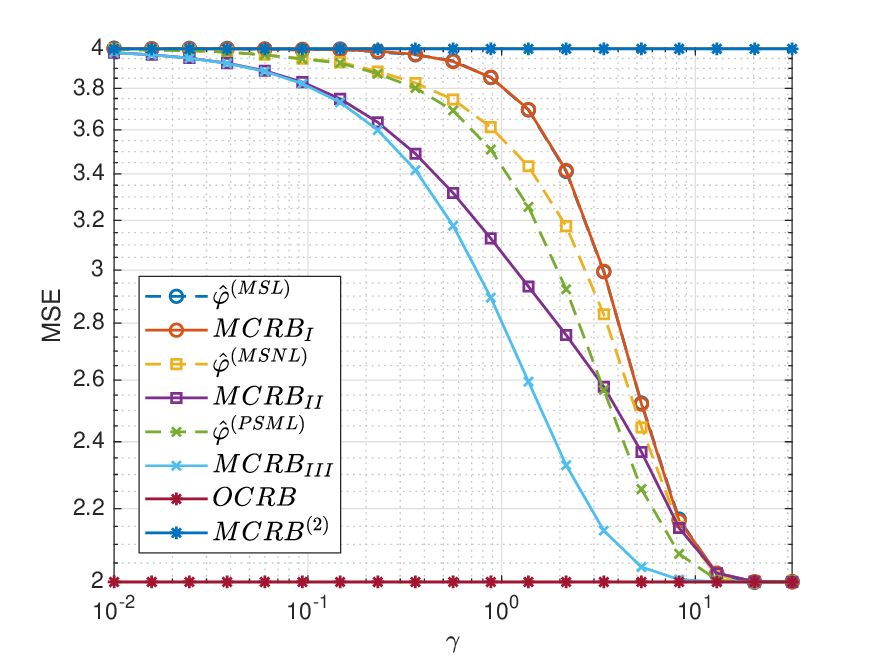}
	 			\vspace{-0.25cm}\caption{ \small{The MSE of the MSL, MSNL, PSML, and the oracle MLs estimators, and PS-MCEBs and the oracle CRB  versus the threshold, $\gamma$, where the true hypothesis is ${\cal H}_1$.}}\vspace{-0.25cm}
	 			\label{MSE_vs_gamma_H0}
	 		\end{figure}

\begin{figure}[htb]
 		\centering \vspace{-0.25cm}
 		\subcaptionbox{ 	\label{H2_bias}}
 		{\includegraphics[width=6.75cm]{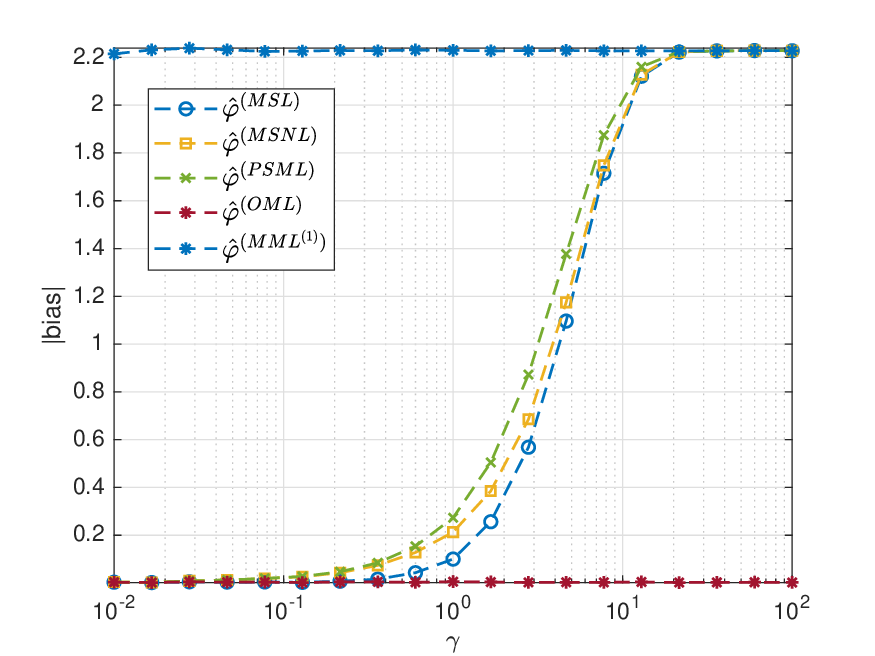}} \vspace{-0.2cm}
 		\subcaptionbox{	\label{H2_MSE}}
 		{\includegraphics[width=6.75cm]{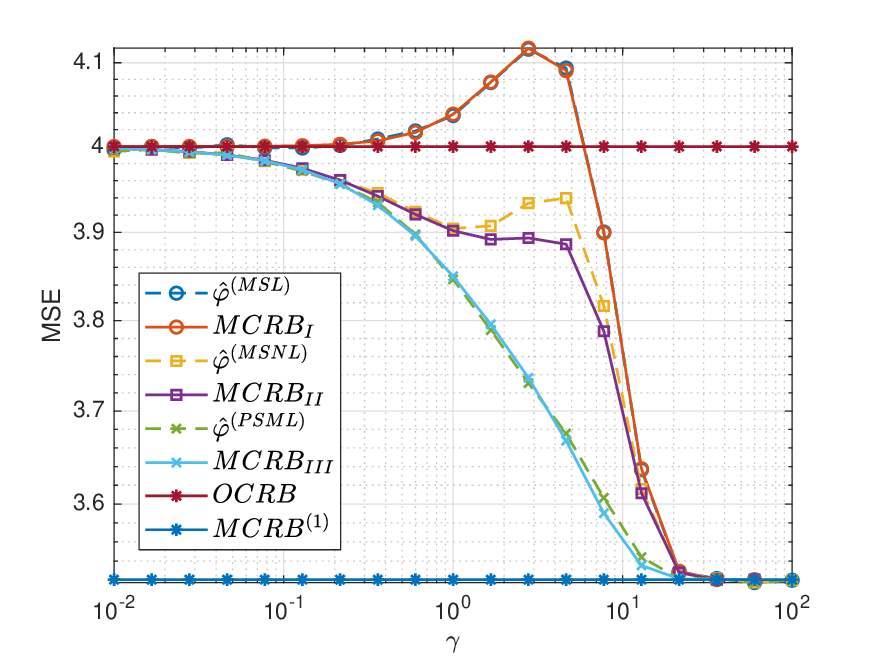}}
 		\caption{\small{The bias (a) and MSE (b) of the MSL, MSNL, PSML, and the oracle MLs estimators,  and PS-MCEBs and the oracle CRB (b) versus the threshold, $\gamma$, where the true hypothesis is ${\cal H}_2$.}}
 		\label{MSE_vs_gamma_H1}\vspace{-0.25cm}
 	\end{figure}

	 In \cref{MSE_vs_gamma_H0,MSE_vs_gamma_H1}, the performance of the MSL, MSNL, and PSML estimators are presented and compared to their corresponding PS-MCRBs and to the oracle CRB, versus the threshold, $\gamma$, where the true hypothesis is ${\cal H}_1$ and ${\cal H}_2$, i.e. where $\bvarphi=\Hmat\thvec^{(1)}$and $\bvarphi=\thvec^{(2)}$ in \cref{MSE_vs_gamma_H0,MSE_vs_gamma_H1}, respectively. In addition,  in \cref{MSE_vs_gamma_H0}, we present the conventional MCRB that always assumes the wrong model (``anti-oracle"), $MCRB^{(2)}$. Similarly, in \cref{H2_bias}, we present the bias of the anti-oracle MML estimator, $\mml{\thvec}^{(1)}$, and in \cref{H2_MSE}, we present conventional MCRB, $MCRB^{(1)}$. The elements of $\Hmat \in \mathbb{R}^{4\times 2}$, $\thvec^{(1)}$, and $\thvec^{(2)}$ were generated once according to a standard Gaussian distribution. The performance is evaluated via $10^6$ Monte-Carlo simulations for $\sigma^2=1$. The estimators and bounds appear in dashed and continuous lines, respectively. 
	 
	Figures  \ref{MSE_vs_gamma_H0} and \ref{H2_MSE} show that in terms of the trace of the MSE matrix from \cref{MSE}, the PSML estimator outperforms the MSNL estimator, and both outperform the commonly-used MSL estimator. In addition, it can be seen that the proposed PS-MCRB for each interpretation is a valid bound that is more informative than the oracle CRB. 
	In \cref{H2_bias}, the $\ell_1$ norm of the bias, 
	$
	\norm{\mathrm{E}_p[\hat{\bvarphi}-\bvarphi]}_1=\sum_{n=1}^N |\mathrm{E}_p[\hat{\varphi}_n-\varphi_n]|,
	$
	is presented for the case where the true hypothesis is ${\cal H}_2$. 
	In this case, the estimators are biased in the conventional sense, since there is a probability of wrong selection that implies bias for any practical estimator. As a result, the oracle CRB is not a valid bound for these estimators, as can be seen in \cref{H2_MSE}. In the case where the true hypothesis is ${\cal H}_1$, the bias of all the estimators is negligible,  and, thus, is not shown here.
	
	In both cases, for the smallest value of $\gamma$, $p_1 \approx 1$, and for the largest value $p_2 \approx 1$. At these extreme points, in practice, only one candidate model is selected.
	Thus, in \cref{MSE_vs_gamma_H0}, for the smallest $\gamma$, ${\cal H}_2$ (in this case, the wrong model) is selected a.s., and, thus, all the estimators and bounds coincide with the conventional MCRB with $f_2(\xvec;\thvec^{(2)})$ as the assumed pdf. For the largest value of $\gamma$, ${\cal H}_1$ (the true model) is selected a.s., and thus, all the estimators and bounds coincide with the oracle ML and oracle CRB.
 Similarly, in \cref{MSE_vs_gamma_H1}, for the smallest $\gamma$, ${\cal H}_2$ (the true model) is selected a.s. and all estimators coincide with the oracle ML estimator, and the MSE of all the estimators and bounds coincide with the oracle CRB.
  For the largest $\gamma$, ${\cal H}_1$ is (wrongly) selected a.s., and the biases of all the estimators coincide with the bias of the conventional MML that takes $f_1(\xvec;\thvec^{(1)})$ as the assumed pdf. Thus, all the estimators and bounds coincide with the conventional MCRB. 
  
	\section{Conclusions}\label{conclusions}
	 In this paper, we investigate the framework of non-Bayesian estimation
after model selection, where we address this problem 
as an estimation under model misspecification. 
We show that the naive interpretation of this scheme results in a non-valid assumed pdf, while the straightforward pdf-corrected normalized interpretation is valid, but creates an incoherent coupling between the parameters of the different candidate models. Thus, we propose 
 the selective inference interpretation, which uses conditional likelihoods, given the event that a particular model was selected. This selective interpretation is valid, coherent, and has appealing properties.
 Based on the three interpretations, we derive the associated MML estimators: the MSL, MSNL, and PSML estimators. 
  In addition, we develop the post-model-selection pseudo-true parameter vectors, the post-model-selection MSE, and the post-model-selection
  unbiasedness.
We propose the PS-MCRB, a novel lower bound on any post-model-selection unbiased estimator that
 incorporates the misspecification in the form of a model-selection procedure.
     We derive the PS-MCRBs for the three interpretations and analyze their properties.
	Finally, we demonstrate in simulations the relations between the proposed estimators and bounds.
	We show that in terms of MSE, the PSML and MSNL estimators, associated with the normalized and selective inference interpretation, outperform the commonly-used MSL estimator, which is associated with the naive interpretation.
	 In addition, we show that the proposed  PS-MCRBs under the different interpretations are more informative than the oracle CRB, where the selective inference interpretation results in the lowest bound and MSE.

 \vspace{-0.1cm}
	\FloatBarrier 
		 \nocite{}
	\bibliographystyle{IEEEtran}
	\bibliography{post_model_selection_bounds_bib} 
\end{document}


\maketitle

	In this report, we present comprehensive development of estimators and bounds for the example section in [\cref{simulations}, \cite{harel2023}]. All the definitions and notations are presented in \cite{harel2023}. 
	\section{Example: Estimation after Channel Detection} \label{example_supp}
	 
In  \cite[\cref{simulations}]{harel2023}, $\xvec \in \mathbb{R}^N$ is assumed to be an observation vector such that (see \cref{model_example})
	 $\xvec =\bvarphi +\wvec$,
	 where $\bvarphi \in \mathbb{R}^N$ is an unknown deterministic parameter vector to be estimated, and $\wvec$ is a white Gaussian noise with zero mean and a known covariance matrix, $\sigma^2\Imat$.
	The two hypotheses regarding $\bvarphi$ are given in \cref{hypotheses}.
Thus, it can be seen that	 in the considered setting,  the candidate pdfs are both Gaussian, with means $\Hmat\thvec^{(1)}$ and $\thvec^{(2)}$, respectively, i.e.
	 \be \label{f_k_Gaussian}
	 f_k(\xvec;\thvec^{(k)})=\begin{cases}
	 	(2\pi\sigma^2)^{-\frac{N}{2}}\exp\left(-\frac{1}{2\sigma^2}\norm{\xvec-\Hmat\thvec^{(1)}}^2\right), &k=1 \\
	 	(2\pi\sigma^2)^{-\frac{N}{2}}\exp\left(-\frac{1}{2\sigma^2}\norm{\xvec-\thvec^{(2)}}^2\right), &k=2.
	 \end{cases}
	 \ee
  In addition, under the considered GLRT selection rule from \cref{Psi_GLRT2},
	the true and the assumed probabilities of selection are given by \cref{pk_chi2}
	and \cref{pi_k_chi2}, respectively.
	In particular, for the considered setting, $\pi_1(\thvec^{(1)})$ is not a function of the parameter $\thvec^{(1)}$. Hence, in the following, we use the notation, $\pi_1(\thvec^{(1)})=\pi_1$.
	 \subsection{Estimators}
	 In this subsection, we derive the estimators from  \cref{supp_estimators} in \cite{harel2023}.
	 \subsubsection{Naive Interpretation-  MSL Estimator}\label{supp_MSL_estimator}
	 The MSL estimator from \cref{MSL_estimator_section} is given by the maximization in \cref{MSL_2}, which can be obtained via
	 the solution to the following score equation
	 \be \label{score_I}
	 	\grad{\thvec^{(k)}}\log f_k(\xvec;\thvec^{(k)})=\zerovec, ~~~ \forall \xvec \in \Am{k}.
	 \ee
	 The gradient of \cref{f_k_Gaussian} w.r.t. $\thvec^{(k)}$  is given by
	 \be \label{grad_log_fk}
	 \grad{\thvec^{(k)}}\log f_k(\xvec;\thvec^{(k)})=\begin{cases}
	 	\frac{1}{\sigma^2}\Hmat^T(\xvec-\Hmat\thvec^{(1)}), &k=1 \\
	 	\frac{1}{\sigma^2}(\xvec-\thvec^{(2)}), &k=2.
	 \end{cases}
	 \ee

	 \subsubsection{Normalized Interpretation-  MSNL estimator}\label{supp_MSNL_estimator}
        By substituting \cref{pi_k_chi2} in \cref{alpha}, we obtain that for the considered setting, the normalization factor is given by
        \be \label{alpha_example}
            \alpha(\thvec)=\pi_1+ \pi_2(\thvec^{(2)})=F_r( \gamma; 0)+1-F_r\left(\gamma;  \lambda^{(2)}  \right).
        \ee
	 The MSNL estimator is given by the maximization in \cref{MSNL_est_k}, i.e. as
	 the solution of the following score equation
	\be \label{score_II}
	\grad{\thvec^{(k)}}\log f_k(\xvec;\thvec^{(k)})-\grad{\thvec^{(k)}}\log \alpha_k(\thvec^{(k)})=\zerovec
 ,~ \forall \xvec \in \Am{k}.
	\ee
	Since 
$\alpha(\thvec)$ in \cref{alpha_example} is not a function of $\thvec^{(1)}$,  $\alpha_1(\thvec^{(1)})$ from \cref{alpha_k} is not a function of $\thvec^{(1)}$, and thus, under $k=1$, \cref{score_II} is reduced to the same score equation as in \cref{score_I}.
Therefore, for $k=1$ (i.e. $\xvec  \in \Am{1}$) we obtain \cref{MSNL_est}.
On the other hand, for $\xvec  \in \Am{2}$ 
	the MSNL estimator of $\thvec^{(2)}$ is obtained by the maximization in \cref{score_II} w.r.t. $\thvec^{(2)}$. 
Since $\pi_1$ is not a function of $\thvec^{(1)}$ then $\alpha_2(\thvec^{(2)})$ is identical to $\alpha(\thvec)$ is \cref{alpha_example}. Therefore, \cref{score_II} is given by
\be \label{score_II_explicit_2}
 \grad{\thvec^{(2)}}\log f_k(\xvec;\thvec^{(2)})-\grad{\thvec^{(2)}}\log \left(1- F_r( \gamma; \lambda^{(2)}) +  F_r( \gamma; 0)\right)=\zerovec
 ,
 \ee

		In order to derive the last term in \cref{score_II_explicit_2}, we note that the cdf of non-central $\chi^2$ with $r$ degrees of freedom, $F_r(\gamma;\lambda)$, is given by
	\be \label{chi2cdf}
	F_r\left(\gamma,  \lambda  \right)= \sum_{m=0}^{\infty} e^{-\frac{\lambda}{2}} \frac{\left(\frac{\lambda}{2}\right)^m}{m!} F_{r+2m}\left(\gamma,  0  \right).
	\ee
	The derivative of \cref{chi2cdf} w.r.t. the non-centrality parameter, $\lambda$,  is given by
	\cite{cohen1988noncentral}
	\be \label{d_chi2cdf_d_lambda} \begin{aligned}[b]
		\frac{\partial F_r\left(\gamma,  \lambda  \right)}{\partial \lambda}=&-\frac{1}{2}F_r\left(\gamma,  \lambda  \right)+\frac{1}{2}
		\sum_{m=1}^{\infty} e^{-\frac{\lambda}{2}} \frac{\left(\frac{\lambda}{2}\right)^{m-1}}{(m-1)!} F_{r+2m}\left(\gamma,  0  \right)
		\\ &=-\frac{1}{2}F_r\left(\gamma,  \lambda  \right)+\frac{1}{2}\sum_{l=0}^{\infty} e^{-\frac{\lambda}{2}} \frac{\left(\frac{\lambda}{2}\right)^{l}}{(l)!} F_{r+2+2l}\left(\gamma,  0  \right)
		=-\frac{1}{2}F_r\left(\gamma,  \lambda  \right)+\frac{1}{2}F_{r+2}\left(\gamma,  \lambda  \right).
	\end{aligned}
	\ee
	By applying the chain rule, and using
 \cref{lambda_2} and $	\pi_2(\thvec^{(2)})=1- F_r( \gamma; \lambda^{(2)})$ from \cref{pi_k_chi2},
we obtain
	\be \label{grad_pi_2}
	\grad{\thvec^{(2)}}\pi_2(\thvec^{(2)})=\frac{\partial \pi_2(\thvec^{(2)})}{\partial \lambda^{(2)}}\grad{\thvec^{(2)}}\lambda^{(2)}= \frac{1}{\sigma^2}(F_r(\gamma,  \lambda^{(2)})-F_{r+2}(\gamma,  \lambda^{(2)}))\Pmat_\Hmat^\bot\thvec^{(2)}.
	\ee
	Hence, \cref{grad_pi_2} implies that 
  the gradient of the right term on the l.h.s. of \cref{score_II_explicit_2}
 w.r.t. $\thvec^{(2)}$ is given by
	\be \label{grad_log_alpha}
	\grad{\thvec^{(2)}}
\log \alpha(\thvec)=\frac{\grad{\thvec^{(2)}} \pi_2(\thvec^{(2)})}{\alpha(\thvec)}=
	\frac{1}{\sigma^2}\frac{F_r(\gamma,  \lambda^{(2)})-F_{r+2}(\gamma,  \lambda^{(2)}) }{F_r( \gamma; 0)+1-F_r(\gamma;  \lambda^{(2)}  )}\Pmat_\Hmat^\bot\thvec^{(2)}.
	\ee
 By substituting \cref{grad_log_fk,grad_log_alpha} in \cref{score_II_explicit_2}, the score equation of the MSL estimator for $\xvec\in \Am{2}$  is reduced to 
	\cref{MSNL_score}.


	\subsubsection{Selective Inference Interpretation-  PSML estimator}\label{supp_PSML_estimator}
	The PSML estimator  is given by the maximization in \cref{PSML_est_k_log}, i.e. as
	the solution to the following score equation
	\be \label{score_III}
	\grad{\thvec^{(k)}}\log f_k(\xvec;\thvec^{(k)})-\grad{\thvec^{(k)}}\log\pi_k(\thvec^{(k)})=\zerovec, ~~~ \forall \xvec \in \Am{k}.
	\ee
Since 
$\pi_1$ from \cref{pi_k_chi2} is not a function of $\thvec^{(1)}$,   under $k=1$, \cref{score_III} is reduced to the same score equation as in \cref{score_I}.
Therefore, for $k=1$ (i.e. $\xvec  \in \Am{1}$) we obtain \cref{PSML_est0}.

For $\xvec  \in \Am{2}$, the PSML estimator of $\thvec^{(2)}$ is obtained by solving \cref{score_III}. By using \cref{grad_pi_2}, we can verify that
\be \label{grad_log_pi2}
	\grad{\thvec^{(2)}}
\log \pi_2(\thvec^{(2)})=\frac{\grad{\thvec^{(2)}} \pi_2(\thvec^{(2)})}{\pi_2(\thvec^{(2)})}=
	\frac{1}{\sigma^2}\frac{F_r(\gamma,  \lambda^{(2)})-F_{r+2}(\gamma,  \lambda^{(2)}) }{1-F_r(\gamma;  \lambda^{(2)}  )}\Pmat_\Hmat^\bot\thvec^{(2)}.
	\ee
By substituting 
 \cref{grad_log_fk,grad_log_pi2} in \cref{score_III},
	the score equation of the PSML estimator for $\xvec\in \Am{2}$  is reduced to \cref{PSML_score}.

		 \subsection{Pseudo-true Parameters}\label{supp_PT}
	In this subsection, we develop the pseudo-true parameter vectors from \cref{pseudo_true_section} in \cite{harel2023} according to each interpretation.
	\subsubsection{Naive Interpretation} \label{supp_PT_naive}
	The $k$th pseudo-true parameter vector under the naive interpretation is given by the maximization in \cref{k_pt_I}, i.e. as the solution of
 \be \label{PT_I_score}
    \grad{\thvec^{(k)}}\mathrm{E}_p \left[ \log f_k(\xvec;\thvec^{(k)})|\Psi=k \right]=\zerovec, ~~\forall k \in \{1,2\}
 \ee
 
	Since the true pdf, $p(\xvec;\bvarphi)$ is not a function of the parameters $\thvec^{(k)}$, by using \cref{grad_log_fk} we obtain that
	\be \label{expectation_grad_log_fk}
	\grad{\thvec^{(k)}}\mathrm{E}_p \left[ \log f_k(\xvec;\thvec^{(k)})|\Psi=k \right]=\begin{cases}
		\frac{1}{\sigma^2}\Hmat^T(\muvec^{(1)}-\Hmat\thvec^{(1)}), &k=1 \\
		\frac{1}{\sigma^2}(\muvec^{(2)}-\thvec^{(2)}), &k=2.
	\end{cases}
	\ee
	where
$
	\muvec^{(k)}$, 
 $k =1,2$
	is the conditional expectation of $\xvec$ given the selection, as defined in \eqref{mu_k}. In \cref{cond_expectations} of this report, we derive a closed-form expression for $\muvec^{(k)}$, $k=1,2$.
	
	Setting the gradient from \cref{expectation_grad_log_fk} to zero results in \cref{vecvar_I}, i.e. in the following term:
     \be \label{vecvar_I_sup}
    	 	 {\thvecvar}^{(k)}_{I}
    	 	 =\begin{cases}
    	 	 	(\Hmat^T\Hmat)^{-1}\Hmat^T \muvec^{(1)}, &k=1 \\
    	 	 	\muvec^{(2)}, &k=2.
    	 	 \end{cases}
    	 	 \ee
 	
	\subsubsection{Normalized Interpretation}\label{supp_PT_normalized}
	The pseudo-true parameter vector under the normalized interpretation is given by the maximization in \cref{k_pt_II_final}.
	Therefore, it is given by the solution of the following equation
	\be \label{PT_II_score}
			\grad{\thvec^{(k)}}\mathrm{E}_p \left[ \log f_k(\xvec;\thvec^{(k)})|\Psi=k \right]-\grad{\thvec^{(k)}}\log \alpha_k (\thvec^{(k)})=\zerovec.
	\ee
	In the considered setting $\alpha(\thvec)$ in \cref{alpha_example} is not a function of $\thvec^{(1)}$, then $\alpha_k$ is also not a function of $\thvec^{(1)}$ (for both $k=1,2$). Thus, for $k=1$ the solution of  \cref{PT_II_score} is given by
 \be 
	{\thvecvar}^{(1)}_{II}={\thvecvar}^{(1)}_{I}=(\Hmat^T\Hmat)^{-1}\Hmat^T \muvec^{(1)}.
	\ee

	By using \cref{expectation_grad_log_fk,grad_log_alpha}, ${\thvecvar}^{(2)}_{II}$ is given as  the solution of the following equation 
	\be 
	\muvec^{(2)}-\thvec^{(2)}- \frac{F_r(\gamma; \lambda^{(2)} )-F_{r+2}(\gamma; \lambda^{(2)} ) }{F_r( \gamma; 0)+1-F_r\left(\gamma;  \lambda^{(2)}  \right)}\Pmat_\Hmat^\bot\thvec^{(2)}=\zerovec.
	\ee
	
	\subsubsection{Selective Inference Interpretation} \label{supp_PT_selective}
		The pseudo-true parameter vector under the selective inference interpretation is given by the maximization in \cref{k_pt_III}.
		Thus, it is given by the solution of the following equation
		\be \label{score_III_PT}
			\grad{\thvec^{(k)}}\mathrm{E}_p\left[ \log f_k(\xvec;\thvec^{(k)})|\Psi=k \right]-\grad{\thvec^{(k)}}\log \pi_k(\thvec^{(k)})=\zerovec.
		\ee
	Similarly to the normalized interpretation, since $\pi_1$ is not a function of $\thvec^{(1)}$ we obtain that (see also \cref{vecvar_1_III})
	\be 
	{\thvecvar}^{(1)}_{III}={\thvecvar}^{(1)}_{I}=(\Hmat^T\Hmat)^{-1}\Hmat^T \muvec^{(1)}.
	\ee
 By substituting \cref{expectation_grad_log_fk,grad_log_pi2} in \cref{score_III_PT}, ${\thvecvar}^{(2)}_{III}$ is given by the solution of the following equation.
	\be  \label{score_III_PT2}
	\muvec^{(2)}-\thvec^{(2)}- \frac{F_r(\gamma; \lambda^{(2)} )-F_{r+2}(\gamma; \lambda^{(2)} )}{1-F_r(\gamma; \lambda^{(2)} )}\Pmat_\Hmat^\bot\thvec^{(2)}=\zerovec.
	\ee
 In \cref{cond_expectation} in this report we show that 
 \be
    \muvec^{(k)}=  \bvarphi
 				 +(-1)^k\frac{1}{p_k} \left(F_r( \gamma; \lambda)-F_{r+2}( \gamma; \lambda) \right)\Pmat_\Hmat^\bot\bvarphi.
 \ee
	Hence, by substituting $\muvec^{(2)}$ in \cref{score_III_PT2}, we can verify that the solution of \cref{score_III_PT2} is obtained for ${\thvecvar}^{(2)}_{III}=\bvarphi$ as appears in
\cref{vecvar_III}.
	
	 \subsection{Post-selection PS-MCRB} \label{supp_PS_MCRB}
	 	 	  In this subsection, we derive the different PS-MCRBs from \cite[\cref{bounds}]{harel2023}, according to each interpretation. In particular we will derive the matrices $\Amat_{i}^{(k)}(\thvecvar^{(k)}_{i}),\Bmat_{i}^{(k)}(\thvecvar^{(k)}_{i})$, $i \in \{I,II,III\}$.
	
	 	 	  	\subsubsection{Naive Interpretation} \label{supp_PSMCRB_I}
	 	 	  By using \cref{f_k_Gaussian}, the Hessian matrix of the log-likelihood under each of the hypotheses is given by
	 	the matrix in \cref{example1_J_A},
	 	 	  which is not a function of $\thvec^{(k)}$ for any $k$. Substitution  \cref{example1_J_A} in \cref{Ak_I} results in $\Amat_{I}^{(k)}(\thvecvar^{(k)}_{I})$ in 
	 	 	 \cref{example_Ak_I}.

	 	 	   The $k$th  outer-product form information matrix is given by \cref{Bk_I}.
	 	 	  By substituting \cref{grad_log_fk} in \cref{Bk_I} we obtain that
	 	 	  \be \label{outer_prod_1_gen}\begin{aligned}[b]
	 	 	  	\Bmat^{(1)}_I(\thvec^{(1)})
	 	 	  	=\mathrm{E}_p\left[ \frac{1}{\sigma^4}\Hmat^T(\xvec-\Hmat\thvec^{(1)})(\xvec-\Hmat\thvec^{(1)})^T\Hmat  | \Psi=1 \right]= \frac{1}{\sigma^4}\Hmat^T\mathrm{E}_p\left[ (\xvec-\Hmat\thvec^{(1)})(\xvec-\Hmat\thvec^{(1)})^T | \Psi=1 \right] \Hmat, 
	 	 	  \end{aligned}
	 	 	  \ee
	 	 	  and respectively,
	 	 	   \be \label{outer_prod_2_gen}\begin{aligned}[b]
	 	 	  	\Bmat^{(2)}_I(\thvec^{(1)})
	 	 	  	=\mathrm{E}_p\left[ \frac{1}{\sigma^4}(\xvec-\thvec^{(2)})(\xvec-\thvec^{(2)})^T  | \Psi=2 \right] .
	 	 	  \end{aligned}
	 	 	  \ee
	 	 	  We can rewrite the inner expectation on the l.h.s. of \cref{outer_prod_1_gen} as follows
	 	 	  \be \label{cond_cov1}\begin{aligned}[b]
	 	 	  	\mathrm{E}&[(\xvec-\Hmat\thvec^{(1)})(\xvec-\Hmat\thvec^{(1)})^T|\Psi=1]  =  
	 	 	  	\mathrm{E}[(\xvec-\muvec^{(1)}+\muvec^{(1)}-\Hmat\thvec^{(1)})(\xvec-\muvec^{(1)}+\muvec^{(1)}-\Hmat\thvec^{(1)})^T|\Psi=1] 
	 	 	 	\\& 	=\bsigma_{\xvec}^{(1)}+(\muvec^{(1)}-\Hmat\thvec^{(1)})(\muvec^{(1)}-\Hmat\thvec^{(1)})^T,
	 	 	  \end{aligned}
	 	 	  \ee
	 	 	  where 
	 	 	  	$\bsigma_{\xvec}^{(k)}$, defined in \eqref{cov_k}, 
	 	 	   is the conditional covariance matrix of $\xvec$ given $\Psi=k$ w.r.t. the true pdf. A closed-form expression for $\bsigma_{\xvec}^{(k)}$ appears in \cref{sig_k} in the following.
	 	 	  Substitution of \cref{cond_cov1} in \cref{outer_prod_1_gen} results in 
	 	 	  \be \label{outer_prod_1}\begin{aligned}[b]
	 	 	  	\Bmat^{(1)}_I(\thvec^{(1)})
	 	 	  	= \frac{1}{\sigma^4}\Hmat^T\bsigma_{\xvec}^{(1)}\Hmat^T + \frac{1}{\sigma^4}\Hmat^T(\muvec^{(1)}-\Hmat\thvec^{(1)})(\muvec^{(1)}-\Hmat\thvec^{(1)})^T\Hmat^T.
	 	 	  \end{aligned}
	 	 	  \ee
	 	 	The  pseudo-true parameter vector, according to the naive interpretation, $\thvecvar^{(1)}_{I}$, satisfies \cref{PT_I_score}, which implies that 
	 	 	  \be \label{pt_I_a}
	 	 	  \muvec^{(1)}-\Hmat\thvecvar^{(1)}_{I}=\muvec^{(1)}-\Hmat(\Hmat^T\Hmat)^{-1}\Hmat^T\muvec^{(1)}=\Pmat_\Hmat^\bot\muvec^{(1)}.
	 	 	  \ee 
	 	 	  By substituting \cref{pt_I_a} in \cref{outer_prod_1} and since $\Pmat_\Hmat^\bot $ is the orthogonal projection to $\Hmat $, 
	 	 	  the naive interpretation outer-product form information matrix, \cref{outer_prod_1} evaluated at $\thvec^{(1)}=\thvecvar^{(1)}_{I}$ is given by
	 	 	  \be \label{BI1}
	 	 	  	\Bmat^{(1)}_I(\thvecvar^{(1)}_{I})
	 	 	  	= \frac{1}{\sigma^4}\Hmat^T\bsigma_{\xvec}^{(1)}\Hmat.
	 	 	  \ee
	 	 	  
	 	 	  In a similar manner to \cref{cond_cov1}, we can write
	 	 	   \be \label{cond_cov2}\begin{aligned}[b]
	 	 	   	\mathrm{E}&[(\xvec-\thvec^{(2)})(\xvec-\thvec^{(2)})^T|\Psi=2] 
	 	 	   	=\bsigma_{\xvec}^{(2)}+(\muvec^{(2)}-\thvec^{(2)})(\muvec^{(2)}-\thvec^{(2)})^T.
	 	 	   \end{aligned}
	 	 	   \ee
	 	 	  Substitution of \cref{cond_cov2} in \cref{outer_prod_2_gen} results in
	 	 	   \be \label{outer_prod_2}\begin{aligned}[b]
	 	 	  	\Bmat^{(2)}_I(\thvec^{(2)})
	 	 	  	= \frac{1}{\sigma^4}\bsigma_{\xvec}^{(2)} + \frac{1}{\sigma^4}(\muvec^{(2)}-\thvec^{(2)})(\muvec^{(2)}-\thvec^{(2)})^T.	 	 	  \end{aligned}
	 	 	  \ee
	 	 By substituting  \cref{vecvar_I_sup} in \eqref{outer_prod_2}, the naive interpretation outer-product form information matrix is given by
	 	 	   \be \label{BI1}
	 	 	  \Bmat^{(2)}_I(\thvecvar^{(2)}_{I})
	 	 	  = \frac{1}{\sigma^4}\bsigma_{\xvec}^{(2)}.
	 	 	  \ee
	 			\subsubsection{Normalized Interpretation} \label{supp_PSMCRB_II}
	 			In the considered setting, $\alpha(\thvec)$ is not a function of $\thvec^{(1)}$. Thus, the right term on the r.h.s of  \cref{Ak_II} vanishes, and the Hessian form of the information matrix for the normalized interpretation is given by
	 				\be \label{A1_II}\begin{aligned}[b]
	 				\Amat^{(1)}_{II}(\thvecvar_{II}) 
	 				= -\frac{1}{\sigma^2}\Hmat^T\Hmat.
	 			\end{aligned}
	 			\ee
	 			In order to derive $\Amat^{(2)}_{II}(\thvec_{II})$, we first compute the Hessian of $\log \alpha(\thvec)$, which satisfies
	 			 \be \label{Hessian_alpha}\begin{aligned}[b]
	 				\grad{\thvec^{(2)}} ^2 \log \alpha(\thvec)
	 				=\frac{\grad{\thvec^{(2)}}^2\pi_2(\thvec^{(2)})}{\alpha(\thvec)}-\grad{\thvec^{(2)}}  \log \alpha(\thvec)\grad{\thvec^{(2)}} ^T \log \alpha(\thvec),
	 			\end{aligned}
	 			\ee
     where the equality is obtained by using the chain rule and the definition of $\alpha(\thvec)$.
	 			By using the chain rule on the left term on the r.h.s. of \cref{grad_log_alpha}, we obtain 
	 			 \be  \label{Hessian_pi2_raw}  
	 				\grad{\thvec^{(2)}}^2 \pi_2(\thvec^{(2)})=  
	 				\frac{\partial \pi_2(\thvec^{(2)})}{\partial \lambda^{(2)}}\grad{\thvec^{(2)}}^2\lambda^{(2)}+
	 				\frac{\partial^2 \pi_2(\lambda^{(2)})}{\partial (\lambda^{(2)})^2}\grad{\thvec^{(2)}}\lambda^{(2)}\grad{\thvec^{(2)}}^T\lambda^{(2)} . 
	 			\ee
	 			In addition,  the derivative of \cref{d_chi2cdf_d_lambda} w.r.t. the non centrality parameter, $\lambda$, is 
	 			\be \label{d2fdlambda2}
	 			\frac{\partial^2 F_r\left(\gamma,  \lambda  \right)}{\partial \lambda^2}=-\frac{1}{2}\frac{\partial }{\partial \lambda}\left( F_r\left(\gamma,  \lambda  \right)-F_{r+2}\left(\gamma,  \lambda  \right)   \right)
	 			=\frac{1}{4}\left(  F_r\left(\gamma,  \lambda  \right)-2 F_{r+2}\left(\gamma,  \lambda  \right)+F_{r+4}\left(\gamma,  \lambda  \right)  \right).
	 			\ee
	 		By using the definition of $\lambda^{(2)}$ in \cref{lambda_2} one can verify that
	 			\be \label{Hessian_lambda}
	 				\grad{\thvec^{(2)}}^2\lambda^{(2)}=\frac{2}{\sigma^2} \Pmat_\Hmat^\bot.
	 			\ee
	 			Recall that $	\pi_2(\thvec^{(2)})=1- F_r( \gamma; \lambda^{(2)})$ from \cref{pi_k_chi2}. Thus, substitution of \cref{d2fdlambda2,Hessian_lambda} in \cref{Hessian_pi2_raw} results in 
	 			\be \label{Hessian_pi2}\begin{aligned}[b]
	 					\grad{\thvec^{(2)}}^2 \pi_2(\thvec^{(2)})=\frac{1}{\sigma^2}&(F_r(\gamma,  \lambda^{(2)})-F_{r+2}(\gamma,  \lambda^{(2)}))\Pmat_\Hmat^\bot
	 					\\&-
	 				\frac{1}{\sigma^4}(F_r(\gamma,  \lambda^{(2)})-2F_{r+2}(\gamma,  \lambda^{(2)})+F_{r+4}(\gamma,  \lambda^{(2)})) \Pmat_\Hmat^\bot\thvec^{(2)}(\thvec^{(2)})^T\Pmat_\Hmat^\bot
	 			\end{aligned}
	 			.
	 			\ee
	 				Substitution of \cref{Hessian_pi2,grad_log_alpha} in \cref{Hessian_alpha}, and then substituting the result in \cref{Ak_II} results in the following closed form expression for \cref{example_Ak_II}:
	 			\be \label{A2_II_final}\begin{aligned}[b]
	 				&\Amat^{(2)}_{II}(\thvecvar^{(2)}_{II}) = 
	 				-\frac{1}{\sigma^2}\left(\Imat+\frac{F_r(\gamma; \lambda^{(2)} )-F_{r+2}(\gamma; \lambda^{(2)} ) }{F_r(\gamma; 0)+1-F_r(\gamma; \lambda^{(2)} )}\Pmat_\Hmat^\bot\right)
	 				\\&+\frac{1}{\sigma^4}\left(\frac{F_r(\gamma, \lambda^{(2)})-2 F_{r+2}(\gamma,  \lambda^{(2)}  )+F_{r+4}(\gamma,  \lambda^{(2)}  )}{F_r(\gamma;0)+1-F_r(\gamma; \lambda^{(2)} )}+\left(   \frac{F_r(\gamma; \lambda^{(2)} )-F_{r+2}(\gamma; \lambda^{(2)} ) }{F_r(\gamma; 0 )+1-F_r(\gamma; \lambda^{(2)} )}\right)^2\right)\Pmat_\Hmat^\bot\thvecvar^{(k)}_{II}(\thvecvar^{(k)}_{II})^T\Pmat_\Hmat^\bot
	 				.
	 			\end{aligned}
	 			\ee

	 			The outer-product form information matrix under the normalized interpretation is given by \cref{Bk_II}.
	 			Since $\alpha(\thvec)$ is not a function of $\thvec^{(1)}$ and since ${\thvecvar}^{(1)}_{II}={\thvecvar}^{(1)}_{I}$, then
	 			\be
	 				\Bmat^{(1)}_{II}(\thvecvar^{(1)}_{II})=\Bmat^{(1)}_{I}(\thvecvar^{(1)}_{I})=\frac{1}{\sigma^4}\Hmat^T\bsigma_{\xvec}^{(1)}\Hmat^T.
	 			\ee
	 			 Substitution of \cref{outer_prod_2} results in
	 			    \be \label{B2II}\begin{aligned}[b]
	 			 	\Bmat^{(2)}_I(\thvec^{(2)})
	 			 	= \frac{1}{\sigma^4}\bsigma_{\xvec}^{(2)} + \frac{1}{\sigma^4}(\muvec^{(2)}-\thvec^{(2)})(\muvec^{(2)}-\thvec^{(2)})^T
	 			 	-\grad{\thvec^{(k)}}\log \alpha(\thvecvar_{II}) \grad{\thvec^{(k)}}^T \log \alpha(\thvecvar_{II}).	 	 	  \end{aligned}
	 			 \ee
	 			Recall that by the definition, the pseudo-true vector, $ $ satisfies \cref{score_II} i.e. $\grad{\thvec^{(2)}}\log \alpha(\thvecvar^{(2)}_{II})=\frac{1}{\sigma^2}\left( \muvec^{(2)}-\thvecvar^{(2)}_{II}  \right)$ and thus
	 			\be \label{B2_II}\begin{aligned}[b]
	 				\Bmat^{(2)}_{II}(\thvecvar^{(k)}_{II})\hspace{-0.1cm}=
	 				\frac{1}{\sigma^4}\bsigma_{\xvec}^{(2)}.
	 			\end{aligned}
	 			\ee

	 			\subsubsection{Selective Inference Interpretation} \label{supp_PSMCRB_III}
	 			The Hessian form of the Information matrix under the normalized interpretation is given by \cite[\cref{Ak_III}]{harel2023}.
	 			In a similar manner to the normalized interpretation, 	since in our settings, $\pi_1$ is not a function of $\thvec^{(1)}$, thus, we obtain 
	 				\cref{example_Ak_III}.
	 			By using the chain rule, we obtain that
	 			\be \label{grad2log_pi}
	 			\grad{\thvec^{(2)}}^2\log \pi_2(\thvec^{(2)})=\frac{\grad{\thvec^{(2)}}^2 \pi_2(\thvec^{(2)})}{\pi_2(\thvec^{(2)})}-\grad{\thvec^{(2)}}\log \pi_2(\thvec^{(2)})\grad{\thvec^{(2)}}^T\log \pi_2(\thvec^{(2)}).
	 			\ee
				Substitution of \cref{Hessian_pi2} in \cref{grad2log_pi}, and then substituting the result in \cref{Ak_III} results in the following closed form expression for \cref{example_Ak_III}:
				\be \label{A2_III_final}\begin{aligned}[b]
					\Amat^{(2)}_{III}&(\thvecvar^{(2)}_{III}) = 
					-\frac{1}{\sigma^2}\left(\Imat+\frac{F_r(\gamma; \lambda )-F_{r+2}(\gamma; \lambda ) }{1-F_r(\gamma; \lambda )}\Pmat_\Hmat^\bot\right)
					\\&+\frac{1}{\sigma^4}\left(\frac{F_r(\gamma, \lambda)-2 F_{r+2}(\gamma,  \lambda )+F_{r+4}(\gamma,  \lambda  )}{1-F_r(\gamma; \lambda)}+\left(   \frac{F_r(\gamma; \lambda )-F_{r+2}(\gamma; \lambda ) }{1-F_r(\gamma; \lambda )}\right)^2\right)\Pmat_\Hmat^\bot\bvarphi\bvarphi^T\Pmat_\Hmat^\bot
					.
				\end{aligned}
				\ee
				
	 			The outer-product form information matrix under the selective inference interpretation is given by \cite[\cref{Bk_III}]{harel2023}.
	 			Similarly to the normalized interpretation, since $\pi_1$ in these settings is not a function of $\thvec^{(1)}$. In addition, since 
	 			$\thvecvar^{(1)}_{III}=\thvecvar^{(1)}_{I}$ we obtain that
	 			\be
	 				\Bmat^{(1)}_{III}(\thvecvar^{(1)}_{III})\hspace{-0.1cm}=\Bmat^{(1)}_I(\thvecvar^{(1)}_{III})=\frac{1}{\sigma^4}\Hmat^T\bsigma_{\xvec}^{(1)}\Hmat.
	 			\ee
	 			Substitution of \cref{outer_prod_2} results in
	 			\be \label{B2_III_a}\begin{aligned}[b]
	 				\Bmat^{(2)}_{III}(\thvecvar^{(k)}_{III})\hspace{-0.1cm}=
	 				\frac{1}{\sigma^4}\bsigma_{\xvec}^{(2)} + \frac{1}{\sigma^4}(\muvec^{(2)}-\thvecvar^{(k)}_{III})(\muvec^{(2)}-\thvecvar^{(k)}_{III})^T
	 				-\grad{\thvec^{(k)}}\log \pi_2(\thvecvar^{(2)}_{III}) \grad{\thvec^{(2)}}^T \log \pi_2(\thvecvar^{(2)}_{III}).
	 			\end{aligned}
	 			\ee
	 			Since $\thvecvar^{(2)}_{III}$ satisfies \cref{score_III_PT}, \cref{B2_III_a} results in
	 			\be \label{B2_III}\begin{aligned}[b]
	 				\Bmat^{(2)}_{III}(\thvecvar^{(k)}_{III})\hspace{-0.1cm}=
	 				\frac{1}{\sigma^4}\bsigma_{\xvec}^{(2)}.
	 			\end{aligned}
	 			\ee
	 			To conclude, we obtain that the outer-product form of the $k$th post-model-selection information matrices are identical for all the interpretations and are given by \cref{outer_FIM_example}.
	 			\section{Conditional Expectations and Covariance Matrices} \label{cond_expectations}
	 			In this section, we derive analytic, closed forms, expressions for the conditional expectations, $\muvec^{(k)}$, and covariance matrices, $\bsigma_{\xvec}^{(k)}$, from \cref{mu_k,cov_k}, respectively. 
	 			 To this end, in the following proposition, we derive the moment-generating function (MGF), for this case. 
	 			\begin{proposition}
	 				Let $\xvec \in \mathbb{R}^N$ be an observation vector such that
	 				\be
	 				\xvec =\bvarphi +\wvec,
	 				\ee
	 				where $\bvarphi \in \mathbb{R}^N$ is an unknown deterministic parameter vector to be estimated, and $\wvec$ is a white Gaussian noise with zero mean and a known covariance matrix, $\sigma^2\Imat$, i.e. $\xvec\sim {\cal N}(\bvarphi,\sigma^2\Imat)$.
	 				The MGF of $\xvec$ given $\xvec \in \Am{k}$, $m_k(\tvec)$, is given by
	 				\be \label{mgf}
	 					m_k(\tvec) \triangleq \mathrm{E}_p[\exp(\tvec^T\xvec)| \xvec \in \Am{k}]= \frac{1}{p_k}\exp\left(\tvec^T\bvarphi+ \frac{\sigma^2}{2}\tvec^T\tvec\right)p_{k,\tvec},
	 				\ee
	 				where
	 				\be \label{p_kt}
	 				p_{k,\tvec} \triangleq \int_{\Am{k}}  p(\xvec;\bvarphi +\sigma^2\tvec) \mathrm{d}\xvec, ~~~k \in \{1,2\},
	 				\ee 
	 				is the probability that $\xvec \in \Am{k}$ if $\xvec \sim {\cal N}(\bvarphi +\sigma^2\tvec,\sigma^2\Imat)$. Notice that in particular $p_{k,\zerovec}=p_k$.
	 			\end{proposition}
 				\begin{IEEEproof}
 					By definition the MGF of $\xvec$ conditioned by $ \xvec \in \Am{k}$ is given by
 					\be
 					m_k(\tvec) = \mathrm{E}_p[\exp(\tvec^T\xvec)| \xvec \in \Am{k}]=\int_{\Omega_\xvec} \exp(\tvec^T\xvec)  p(\xvec|\xvec \in \Am{k};\bvarphi ) \mathrm{d}\xvec.
 					\ee	
 					By using Bayes rule we obtain that $p(\xvec|\xvec \in \Am{k};\bvarphi )=\frac{p(\xvec;\bvarphi )\ind{\xvec\in \Am{k}}}{p_k}$, therefore
 					\be
 						m_k(\tvec) = \frac{1}{p_k}	\int_{\Am{k}} \exp(\tvec^T\xvec) p(\xvec;\bvarphi ) \mathrm{d}\xvec.
 					\ee	
 					Substitution of $p(\xvec;\bvarphi )$, a Gaussian pdf, results in 
 					\be \begin{aligned}[b]
 						\frac{1}{p_k}&	\int_{\Am{k}} \exp(\tvec^T\xvec) p(\xvec;\bvarphi ) \mathrm{d}\xvec=
 						\frac{1}{p_k}	\int_{\Am{k}} \exp(\tvec^T\xvec) (2\pi\sigma^2)^{-\frac{N}{2}}\exp\left(-\frac{1}{2\sigma^2}\norm{\xvec-\bvarphi}^2\right) \mathrm{d}\xvec
 						\\&=\frac{1}{p_k}	\int_{\Am{k}}(2\pi\sigma^2)^{-\frac{N}{2}}\exp\left(-\frac{1}{2\sigma^2}\norm{\xvec-\bvarphi}^2+\tvec^T\xvec\right) \mathrm{d}\xvec
 						\\&=\frac{1}{p_k}	\int_{\Am{k}}(2\pi\sigma^2)^{-\frac{N}{2}}\exp\left(-\frac{1}{2\sigma^2}(\xvec^T\xvec-2(\bvarphi+\sigma^2\tvec)^T\xvec+\bvarphi^T\bvarphi)\right) \mathrm{d}\xvec
 						\\&=	\frac{1}{p_k}	\int_{\Am{k}} (2\pi\sigma^2)^{-\frac{N}{2}}\exp\left(-\frac{1}{2\sigma^2}\norm{\xvec-(\bvarphi+\sigma^2\tvec)}^2\right)\exp\left(\tvec^T\bvarphi+ \frac{\sigma^2}{2}\tvec^T\tvec\right) \mathrm{d}\xvec
 							\\&=	\frac{1}{p_k} \exp\left(\tvec^T\bvarphi+ \frac{\sigma^2}{2}\tvec^T\tvec\right)	\int_{\Am{k}} (2\pi\sigma^2)^{-\frac{N}{2}}\exp\left(-\frac{1}{2\sigma^2}\norm{\xvec-(\bvarphi+\sigma^2\tvec)}^2\right) \mathrm{d}\xvec.
 					\end{aligned}\ee 
 					Notice that
 					\be
 						(2\pi\sigma^2)^{-\frac{N}{2}}\exp\left(-\frac{1}{2\sigma^2}\norm{\xvec-(\bvarphi+\sigma^2\tvec)}^2\right)= p(\xvec;(\bvarphi+\sigma^2\tvec))
 					\ee
 					which results in \cref{mgf}.
 				\end{IEEEproof}
 			
 				The moments of $\xvec $ given the $k$th selection are obtained by the derivatives of the $k$th MGF, $m_k(\tvec)$. In particular,
 			\be \label{mu_k_mgf}
 			\muvec^{(k)} = \grad{\tvec}m_k(\tvec)\eval{\tvec=\zerovec}
 			\ee
 			and
 			\be \label{sig_k_mgf}
 			\bsigma_{\xvec}^{(k)}= \grad{\tvec}^2 m_k(\tvec)\eval{\tvec=\zerovec}-\muvec^{(k)}(\muvec^{(k)})^T.
 			\ee
 			In the following, we derive closed forms for \cref{mu_k_mgf,sig_k_mgf}.
 			
 			\subsection{Conditional Expectations} \label{cond_expectation}
 			In the settings in \cref{simulations}, we consider the GLRT selection rule from \cref{Psi_GLRT2}, $\psi(\xvec)\sim \chi^2_r(\lambda)$. 
 			Therefore, $p_{k,\tvec}$ from \cref{p_kt} are given by 	
 			\be \label{p_kt_chi}
 				p_{k,\tvec}=\begin{cases}
 					F_r( \gamma; \lambda_\tvec), &k=1 \\
 					1- F_r( \gamma; \lambda_\tvec), &k=2,
 				\end{cases}
 			\ee		
 			where 
 			\be \label{lambda_t}
 			\lambda_\tvec \triangleq  \frac{(\bvarphi+\sigma^2\tvec)^T\Pmat_\Hmat^\bot(\bvarphi+\sigma^2\tvec)}{\sigma^2}.
 			\ee
 			By substitution of and by using the chain rule, we obtain that
 			\be \label{grad_mk_} \begin{aligned}[b]
 							\grad{\tvec}m_k(\tvec) =	 m_k(\tvec)(\bvarphi+\sigma^2\tvec)
 				+\frac{1}{p_k}\exp\left(\tvec^T\bvarphi+ \frac{\sigma^2}{2}\tvec^T\tvec\right)\grad{\tvec}p_{k,\tvec}.
 			\end{aligned}
 			\ee
 			By substituting $p_{k,\tvec}$ from \cref{p_kt_chi} and by using the derivative from \cref{d_chi2cdf_d_lambda} we obtain that
 			\be \label{grad_pk}
 				\grad{\tvec}p_{k,\tvec}= \frac{\partial p_{k,\tvec}}{\partial \lambda_\tvec}\grad{\tvec}\lambda_\tvec
 				=(-1)^k\left(F_r( \gamma; \lambda_\tvec)-F_{r+2}( \gamma; \lambda_\tvec) \right)\Pmat_\Hmat^\bot(\bvarphi+\sigma^2\tvec).
 			\ee
 			\be \label{grad_mk} 
 				\grad{\tvec}m_k(\tvec) =	 m_k(\tvec)(\bvarphi+\sigma^2\tvec)
 				+(-1)^k\frac{1}{p_k}\exp\left(\tvec^T\bvarphi+ \frac{\sigma^2}{2}\tvec^T\tvec\right)\left(F_r( \gamma; \lambda_\tvec)-F_{r+2}( \gamma; \lambda_\tvec) \right)\Pmat_\Hmat^\bot(\bvarphi+\sigma^2\tvec).
 			\ee
 			Evaluation of \cref{grad_mk} at $\tvec=\zerovec$ results in 
 			\be \label{muvec_k}
 				\muvec^{(k)} = \grad{\tvec}m_k(\tvec)\eval{\tvec=\zerovec}= \bvarphi
 				 +(-1)^k\frac{1}{p_k} \left(F_r( \gamma; \lambda)-F_{r+2}( \gamma; \lambda) \right)\Pmat_\Hmat^\bot\bvarphi.
 			\ee
 			\subsection{Conditional Covariance Matrices}
 			By taking the gradient w.r.t. $\tvec$ on \cref{grad_mk} we obtain that
 			\be \label{Hessian_mk} \begin{aligned}[b]
 				\grad{\tvec}^2m_k(\tvec)= m_k(\tvec)\sigma^2\Imat+ \grad{\tvec}m_k(\tvec)(\bvarphi+\sigma^2\tvec)^T 
 				+\frac{1}{p_k}\exp\left(\tvec^T\bvarphi+ \frac{\sigma^2}{2}\tvec^T\tvec\right)\left((\bvarphi+\sigma^2\tvec)\grad{\tvec}^Tp_{k,\tvec}
 				+\grad{\tvec}^2p_{k,\tvec}\right)
 			\end{aligned}
 			\ee
 			where $\grad{\tvec}p_{k,\tvec}$ is given in \cref{grad_pk} and
 			\be \label{Hessian_pk}\begin{aligned}[b]
 					\grad{\tvec}^2&p_{k,\tvec}
 				=(-1)^{k-1}\left(F_r( \gamma; \lambda_\tvec)-2F_{r+2}( \gamma; \lambda_\tvec) + F_{r+4}( \gamma; \lambda_\tvec)\right)\Pmat_\Hmat^\bot(\bvarphi+\sigma^2\tvec)(\bvarphi+\sigma^2\tvec)^T\Pmat_\Hmat^\bot
 			\\&+(-1)^k\left(F_r( \gamma; \lambda_\tvec)-F_{r+2}( \gamma; \lambda_\tvec) \right)\Pmat_\Hmat^\bot\sigma^2.
 			\end{aligned}
 			\ee
 				Evaluation of \cref{Hessian_mk} at $\tvec=\zerovec$ results in
 					\be \label{second_moment} \begin{aligned}[b]
 					\grad{\tvec}^2&m_k(\tvec)\eval{\tvec=\zerovec}= \sigma^2\Imat+\muvec^{(1)}\bvarphi^T
 					+\frac{1}{p_k}	(-1)^k\left(F_r( \gamma; \lambda)-F_{r+2}( \gamma; \lambda) \right)\bvarphi	\bvarphi^T  \Pmat_\Hmat^\bot
 					\\+&
 					(-1)^{k-1}\frac{1}{p_k}\left(F_r( \gamma; \lambda)-2F_{r+2}( \gamma; \lambda) + F_{r+4}( \gamma; \lambda)\right)\Pmat_\Hmat^\bot\bvarphi\bvarphi^T\Pmat_\Hmat^\bot
 					+(-1)^k\frac{1}{p_k}\left(F_r( \gamma; \lambda)-F_{r+2}( \gamma; \lambda) \right)\Pmat_\Hmat^\bot\sigma^2
 				.\end{aligned}
 				\ee 
 				Therefore, by substituting \cref{muvec_k,second_moment} we obtain that
 				\be \label{sig_k} \begin{aligned}[b]
 						\bsigma_{\xvec}^{(k)}=& \grad{\tvec}^2 m_k(\tvec)\eval{\tvec=\zerovec}-\muvec^{(k)}(\muvec^{(k)})^T
 					=\sigma^2\Imat+
 					(-1)^{k-1}\frac{1}{p_k}\left(F_r( \gamma; \lambda)-2F_{r+2}( \gamma; \lambda) + F_{r+4}( \gamma; \lambda)\right)\Pmat_\Hmat^\bot\bvarphi\bvarphi^T\Pmat_\Hmat^\bot
 					\\&+(-1)^k\frac{1}{p_k}\left(F_r( \gamma; \lambda)-F_{r+2}( \gamma; \lambda) \right)\Pmat_\Hmat^\bot\sigma^2
 					-\left(\frac{1}{p_k}\left(F_r( \gamma; \lambda)-F_{r+2}( \gamma; \lambda) \right) \right)^2\Pmat_\Hmat^\bot\bvarphi\bvarphi^T\Pmat_\Hmat^\bot,
 				\end{aligned}
 				\ee
 				
%

 		\bibliographystyle{IEEEtran}
	\bibliography{post_model_selection_bounds_bib}